\documentclass[fleqn,usenatbib]{mnras}

\usepackage{newtxtext,newtxmath}
\usepackage[T1]{fontenc}

\DeclareRobustCommand{\VAN}[3]{#2}
\let\VANthebibliography\thebibliography
\def\thebibliography{\DeclareRobustCommand{\VAN}[3]{##3}\VANthebibliography}

\usepackage{graphicx}	% Including figure files
\usepackage{amsmath}	% Advanced maths commands
\usepackage[dvipsnames]{xcolor}
\newcommand{\msun}{M$_{\odot}$}	% per cm-squared
\newcommand{\siggas}{$\Sigma_{\rm gas}$}	% per cm-squared
\newcommand{\fy}{$\phi$}

\newcommand{\kmps}{km s$^{-1}$}
\newcommand{\zsol}{$Z_{\odot}$}
\newcommand{\qedi}{\citetalias{QEDI}}

\newcommand{\aref}[1]{\hyperref[#1]{Appendix~\ref{#1}}}
\defcitealias{QEDI}{QED I}
\defcitealias{Huang24a}{QED II}

\title[QED III - Outflows as function of environment ]{QUOKKA-based understanding of outflows (QED) - III. Outflow loading and phase structure as a function of galactic environment}

\author[Vijayan et al.]{
Aditi Vijayan$^{1,2}$\thanks{E-mail:aditi.vijayan@anu.edu.au},
Mark R. Krumholz$^{1,2}$,
Benjamin D. Wibking$^{3}$
\\
$^{1}$Research School of Astronomy and Astrophysics, Australian National University, Canberra ACT 2601, Australia\\
$^{2}$ARC Centre of Excellence for Astronomy in Three Dimensions (ASTRO-3D), Canberra ACT 2601, Australia\\
$^{3}$Department of Physics and Astronomy, Michigan State University, 567 Wilson Road, East Lansing, MI 48824, USA\\
}

\date{Accepted XXX. Received YYY; in original form ZZZ}

\pubyear{2015}

\begin{document}
\label{firstpage}
\pagerange{\pageref{firstpage}--\pageref{lastpage}}
\maketitle

% Abstract of the paper
\begin{abstract}
We present results from a suite of 3D high-resolution hydrodynamic simulations of supernova-driven outflows from galactic disc regions with a range of gas surface density, metallicity, and supernova scale height. We use this suite to quantify how outflow properties -- particularly the loading factors for mass, metallicity, and energy -- vary with these parameters. We find that the winds fall into three broad categories: steady and hot, multiphase and moderately bursty, and cool and highly bursty. The first of these is characterised by efficient metal and energy loading but weak mass loading, the second by moderate loading of mass, metals, and energy, and the third by negligible metal and energy loading but substantial mass loading. The most important factor in determining the kind of wind a galaxy will produce is the ratio of supernova to gas gas scale heights, with the latter set by a combination of supernova rate, metallicity-dependent cooling rate, and the gravitational potential. These often combine in counterintuitive ways -- for example increased cooling causes cold clouds to sink into the galactic midplane more rapidly, lowering the volume-filling factor of dense gas and making the environment more favourable for strong winds. Our findings suggest that the nature of galactic winds is likely highly sensitive to phenomena such as runaway stars occuring at a large height and dense gas and are poorly captured in most simulations, and that metal loading factors for type Ia supernovae may be substantially larger than those for type II, with important implications for galactic chemical evolution.
\end{abstract}

% Select between one and six entries from the list of approved keywords.
% Don't make up new ones.

%%%%%%%%%%%%%%%%%%%%%%%%%%%%%%%%%%%%%%%%%%%%%%%%%%

%%%%%%%%%%%%%%%%% BODY OF PAPER %%%%%%%%%%%%%%%%%%

\section{Introduction}

% 1. This is a follow up study. Summarise what was done in \qedi\ and the main results. 

Feedback from supernovae is a key component of the large scale baryon cycle that drives the evolution of galaxies. In a  Milky Way-mass galaxy, this feedback is likely strong enough to launch gaseous flows capable of carrying substantial amounts of mass and newly-synthesized metals out of the galactic disc and into the circum-galactic medium. However, quantifying the properties of the outflowing gas is a challenging problem owing to its complicated multiphase structure and the resulting very large dynamic range involved in the physical processes responsible for setting these properties. 

Because the properties of the outflowing gas depend critically on the conditions of the interstellar medium (ISM) from which they are launched, ideally we would like to simulate the entire disc of a galaxy and the outflows it generates self-consistently, with enough resolution to capture the Sedov-Taylor phase of supernova expansion and thereby avoid subgrid models, and properly including metal injection by supernovae so that it is possible to study metal loading. The ultimate goal of such studies would be to understand how the mass, energy, and metal loading of galactic winds -- which characterise the ratios of wind mass flux to star formation rate, wind energy flux to supernova mechanical energy release, and wind metal flux to metal synthesis rate, respectively -- vary as a function of the properties of the galaxy that launches the winds. Determining this relationship is a key step toward a complete model of galaxy formation and galactic chemical evolution, because the loading factors play a decisive role in models of these processes. For example, mass loading factors largely determine the stellar mass-halo mass relation and the cosmic star formation history \citep[e.g.,][]{Lilly13a, Dekel14b}, energy loading factors are critical to the energy balance of the circumgalactic medium \citep[e.g.,][]{Suresh15a, Li20b}, and metal loading factor shape both the mass-metallicity relation and the gradients of metallicity within galaxies \citep[e.g.,][]{Peeples&Shankar2011, Forbes19a, Sharda21a, Sharda21b}. While in principle one can attempt to determine the loading factors empirically by comparing models to observations using machine learning or similar approaches \citep[e.g.,][]{Forbes19a, Villaescusa-Navarro21a}, in order to have confidence in the results these methods generate we must complement them with first-principles physical simulations that measure loading factors directly.

Such efforts are (barely) feasible for simulations of small numbers of isolated dwarf galaxies \citep[e.g.,][]{Emerick+2018, Andersson23a, Rey24a, Steinwandel+24, Schneider24a}, but at present such an approach is too expensive to use for parameter studies of larger galaxies, which are required if we are to carry out systematic studies of the relationships between outflows and the different physical quantities that might affect them.
Given this reality, systematic parameter studies have mostly adopted the ``tall box'' approximation, whereby one simulates a portion of a galactic disc by treating it as a periodic box in the directions parallel to the galactic plane, while in the direction orthogonal to the plane one uses a domain that is significantly larger and with some type of outflow boundary condition. Recent examples of this approach include the SILCC \citep[e.g.,][]{Girichidis+16-SILCC, Girichidis+18} and TIGRESS/SMAUG \citep[e.g.][]{Kim&Ostriker2017, Kim+20} simulation suites, along with several earlier efforts \citep[e.g.,][]{Creasy+15,Li&Bryan2017}. Even the most recent of these, however, struggle with resolution and simulation volume when carrying out large parameter studies; for example, SMAUG achieves 2 pc resolution in a $0.5\times 0.5\times 3.6$ kpc$^3$ region, or 4 pc resolution for a $1\times 1\times 7.2$ kpc$^3$ region.

In \citet[hereafter \citetalias{QEDI}]{QEDI}, we introduced a new suite of high resolution tall-box simulations, QED\footnote{\textsc{Quokka}-based Understanding of Outflows Derived from Extensive, Repeated,
Accurate, Thorough, Demanding, Expensive, Memory-consuming,
Ongoing Numerical Simulations of Transport, Removal, Accretion,
Nucleosynthesis, Deposition, and Uplifting of Metals (QUOD ERAT
DEMONSTRANDUM, or QED)}. The QED simulations use the new radiation-hydrodynamics code \textsc{Quokka} \citep{Wibking23a, He24a}, which leverages GPU acceleration to reach combinations of spatial resolution and simulation volume significantly larger than previous efforts. \qedi~presented a simulation of a $1\times1\times 8$ kpc$^3$ patch of a galactic disc with conditions chosen to be similar to those of the Solar Neighbourhood, captured at a uniformly-high resolution of 2 pc -- comparable to the best available resolutions used in previous works, but with nearly an order of magnitude larger volume that allowed us to follow the wind farther from the disc while still capturing the complex phase structure of the outflow. We found that this combination of volume and resolution is sufficient for the mass and metal outflow rates and the metallicity of the outflows for a pristine background to converge (see Figure 6 in \citetalias{QEDI}), and that SN-driven winds are highly loaded with metals. Conversely, we found that a number of earlier simulations were likely converged with respect to bulk mass loading, but not with respect to metal loading, which is more numerically challenging to capture.

Much of the challenge comes from the fact that metals are not uniformly distributed amongst the different temperature phases. They are mostly carried by the hot phase at lower heights where the winds are first launched. As outflows move away from the disc the gap between phases narrows due to the exchange of mass and metals between them, to the extent that, for regions where the initial ISM metallicity is already enriched to Solar levels, the difference in metal loading across phases becomes small by the time the gas reaches $\approx 4$ kpc off the disc. In \citet{Huang24a}, we show that this mixing process leaves observable traces in spatially-resolved X-ray spectra of edge-on galactic winds, and is the likely explanation for the negative metallicity gradient observed in some $\alpha$-elements in the wind of M82 \citep{Lopez+2020}.

Though the Solar Neighbourhood provides a representative case for understanding metal loading, the characteristics of the outflowing gas can change dramatically with environment \citep{Li&Bryan2017, Kim+20}. In this paper, we present a broad parameter study using the QED framework to explore how parameters like the star formation rate and gas surface density (linked via the \citealt{Kennicutt98a} relation), metallicity and thus gas cooling rate, and the vertical distribution of supernovae, might affect the outflow properties. Our goal is to develop a general understanding of the relationship between outflow properties and galactic properties, with a particular focus on metal loading, leveraging the combination of high resolution and high volume made possible by GPU acceleration and that our experiments in \qedi~show are particularly important for this parameter. The outline of the rest of this paper is as follows. In \autoref{sec:methods}, we briefly describe the simulation and analysis setup. We present our results in \autoref{sec:results} and discuss their implications in \autoref{sec:disscussion}.

\section{Methods}\label{sec:methods}

All our simulations use the \textsc{Quokka} code, which solves the Euler equations of compressible gas dynamics together with gravity and radiative cooling. The physics included in our simulations is largely similar to that in \citetalias{QEDI}, so here we focus only on aspects of the simulations that differ from those used in that paper.

\subsection{Model grid}

Our simulations can be described by three parameters, which we vary to generate a grid of models. These parameters are the surface densities of gas, stars, and star formation (which all vary together), the metallicity of the gas, and the scale height over which supernovae are injected. We encode the choice of parameter values for each run in the run name, which takes the form $\Sigma$sss-Zzzz-Hhhh, where `sss' gives the gas surface density in units of M$_\odot$ pc$^{-2}$, `zzz' gives the metallicity relative to Solar, and `hhh' gives the scale height of supernovae in units of pc; details of how we implement these initial conditions and how they affect the physics of the simulation are provided below. Thus for example run $\Sigma$13-Z1-H150 has an initial gas surface density of 13 M$_\odot$ pc$^{-2}$, Solar metallicity, and supernova explosions with a scale height of 150 pc. 

We list the full set of simulations we have carried out and summarise their parameters in \autoref{tab:params}. The motivation for our grid is as follows. The case $\Sigma$13-Z1-H150 represents a set of parameters typical of the Solar neighbourhood, and matches the case presented in \citetalias{QEDI} (although we have re-run the simulation here due to minor differences in the procedure that we detail below). In order to explore the effects of surface density we then carry out runs $\Sigma$50-Z1-H150 and $\Sigma$2.5-Z1-H150, which have gas surface densities four times larger and smaller, respectively. The former case might represent a weak circumnuclear starburst or inner galaxy, while the latter is typical of outer spiral galaxies or diffuse dwarfs.
The metallicity of the gas affects the rate at which it cools, and to explore this parameter we add runs $\Sigma$2.5-Z0.2-H150, $\Sigma$13-Z0.2-H150, and $\Sigma$50-Z0.2-H150, which are identical to the corresponding Z1 runs except that the metallicity is reduced to a value more representative of dwarf galaxies such as the Small Magellanic Cloud. Finally, the run series $\Sigma$2.5-Z1-H300, $\Sigma$2.5-Z1-H1000, $\Sigma$2.5-Z1-H1500, and $\Sigma$2.5-Z1-H3000, together with the run $\Sigma$2.5-Z1-H150, allow us to explore how the vertical distribution of supernovae affects outflow generation.

We note that not all of these cases are likely realised in typical galaxies. For example, circumnuclear starbursts with metallicities as low as $Z=0.2Z_\odot$ are rare, at least in the local Universe, and conversely outer galaxies and dwarfs usually have sub-Solar metallicites. Similarly, while the scale heights of SNe are not perfectly correlated to the scale height of the ISM due to factors like many O stars being runaways and type Ia SNe that occur in older stellar populations with larger scale heights, they are clearly not entirely uncorrelated either. Our motivation for exploring this full grid rather than limiting our exploration to the parts of it most reminiscent of observed galaxies is that doing so gives us the baseline necessary to understand the physics of galactic wind driving.

All our runs take place in a domain of size $1\times 1\times 8$ kpc except for the $\Sigma$2.5 runs, which use domains of size $1\times 1\times 16$ kpc; this larger vertical extent is needed to ensure that there is enough room for a wind-dominated zone to develop in runs where the gas scale height is large because gravity is weak. We use a resolution of 2 pc -- chosen based on the convergence tests reported in \citetalias{QEDI} -- in all runs except the $\Sigma$2.5 ones, for which we use $\Delta x = 4$ pc in order to keep the computational cost reasonable. Thus all runs except the $\Sigma$2.5 series use $512 \times 512 \times 4096$ cells, while the $\Sigma$2.5 runs use $256\times 256\times 4096$ cells. The duration of the runs varies from $208-300$ Myr (as indicated in the last column of \autoref{tab:params}), which in all cases is sufficient for the outflows to reach statistical steady-state.

As in \citetalias{QEDI} we adopt periodic boundaries in the $x$ and $y$ directions. In the $z$ direction we adopt ``diode'' boundary conditions, which is a change from \citetalias{QEDI} in which we used outflow boundary conditions in $z$. The difference is that for diode boundary conditions we adopt the usual outflow boundary condition -- linearly extrapolating all simulation variables into the ghost zones -- only in locations where the computational cell adjacent to the boundary has a velocity vector that points out of the domain. In locations where the adjacent cell has an inward velocity, we instead switch to reflecting boundary conditions, whereby we mirror the values inside the domain into the ghost cells, but reverse the sign of the momentum component normal to the surface. This choice ensures that mass is free to leave the domain, but that no new mass can enter. We implement this new boundary condition because we find that, while outflow and diode boundaries produce very similar results for the initial conditions explored in \citetalias{QEDI}, they lead to quite different outcomes in some of the runs we carry out here. We discuss the implications of this finding further in \autoref{sec:disscussion}.

\begin{table*}
\begin{center}
\begin{tabular}{|l|c|c|c|c|c|c|c|c|c|c|c|c|c|}
\hline
\hline
Name & \siggas & $\Sigma_*$  &  $\sigma_{1}$ & $\rho_{1,0}/m_\mathrm{H}$  & $\rho_{\rm dm}$ & $R_0$ & $\Sigma_{\rm SFR}$ & $h_{\rm SN}$ & $Z_{\rm bg}$ & $\Delta x$  & $L_z$ &  $t_f$  \\
 &  [M$_\odot$ pc$^{-2}$] &  [M$_\odot$ pc$^{-2}$] & [km s$^{-1}$] & [cm$^{-3}$] & [M$_\odot$ pc$^{-3}$] 
 & [kpc] & [M$_{\odot}$ yr$^{-1}$ kpc$^{-2}$] & [pc]  & [\zsol] & [pc] & [kpc] & [Myr] 
 \\
(1)  & (2) & (3) & (4) & (5) & (6)  & (7) & (8) & (9) & (10) & (11) & (12) & (13) \\
\hline
% \multicolumn{9}{c}{Fixed grid runs} \\
\hline
$\Sigma$13-Z1-H150 & $13$   & $42$   &  $7$ & $1.688$ & $6.4\times 10^{-3}$ & $8$ & $6\times 10^{-3}$  & $150$ & $1$ & $2$ & $4$ &  $230$  \\
\\
$\Sigma$50-Z1-H150 & $50$   & $208$   &  $15$  & $6.968$ & $2.4\times 10^{-2}$ & $4$ & $3.9\times 10^{-2}$ & $150$ & $1$ & $2$ & $4$ &  $221$  \\
\\
$\Sigma$2.5-Z1-H250 & $2.5$   & $1.71$   &  $11$  & $0.0268$ & $1.4\times 10^{-3}$ & $16$ & $1.58\times 10^{-4}$ & $150$ & $1$ & $4$ & $8$ & $300$  \\

\\

$\Sigma$13-Z0.2-H150 & $13$   & $42$   &  $7$ & $1.688$ & $6.4\times 10^{-3}$ & $8$ &  $6\times 10^{-3}$ & $150$ & $0.2$ & $2$ & $4$ & $208$ \\
\\

$\Sigma$50-Z0.2-H150 & $50$   & $208$   &  $15$  & $6.968$ & $2.4\times 10^{-2}$ & $4$ & $3.9\times 10^{-2}$ & $150$ & $0.2$ & $2$ & $4$ & $290$ \\
\\
$\Sigma$2.5-Z0.2-H150 & $2.5$   & $1.71$   &  $11$  & $0.0268$ & $1.4\times 10^{-3}$ & $16$ & $1.58\times 10^{-4}$ & $150$ & $0.2$ & $4$ & $8$ & $300$\\

\\

$\Sigma$2.5-Z1-H300 & $2.5$   & $1.71$   &  $11$  & $0.0268$ & $1.4\times 10^{-3}$ & $16$ & $1.58\times 10^{-4}$ & $300$ &$1$ & $4$ & $8$ & $300$ \\
\\
$\Sigma$2.5-Z1-H1000 & $2.5$   & $1.71$   &  $11$  & $0.0268$ & $1.4\times 10^{-3}$ & $16$ & $1.58\times 10^{-4}$ & $1000$ & $1$ & $4$ & $8$ & $300$ \\
\\

$\Sigma$2.5-Z1-H1500 & $2.5$   & $1.71$   &  $11$  & $0.0268$ & $1.4\times 10^{-3}$ & $16$ & $1.58\times 10^{-4}$ & $1500$ &$1$ & $4$ & $8$ &  $300$\\
\\
$\Sigma$2.5-Z1-H2000 & $2.5$   & $1.71$   &  $11$  & $0.0268$ & $1.4\times 10^{-3}$ & $16$ & $1.58\times 10^{-4}$ & $2000$ & $1$ & $4$ & $8$ & $300$ \\

\hline
\hline
\\

\end{tabular}
\caption{Summary of parameters for all runs. Columns 2-7 describe the initial parameters that control the initial density distribution and gravitational potential: gas surface density ($\Sigma_{\rm gas}$), stellar surface density ($\Sigma_*$), dispersion ($\sigma_1$) and midplane density ($\rho_{1,0}$) of the gas component containing most of the mass, dark matter density ($\rho_{\rm dm}$), and Galactocentric radius ($R_0$). Columns 8 and 9 provide the parameters that control feedback: surface density of star formation ($\Sigma_{\rm SFR}$) and the scale-height of SN explosions ($h_{\rm SN}$).
Column 10 lists the metallicity ($Z_{\rm bg}$) which sets the cooling rate of gas. Column 11 is the resolution ($\Delta x$) which is uniform throughout the box. Column 12  is the box half-height ($L_z$). Column 13 gives the total time for which the simulations have been evolved.}
\label{tab:params}
\end{center}
\end{table*}

\subsection{Gravitational potential and initial gas density profile}\label{subsec:gas_grav}

As in \citetalias{QEDI}, we do not include gas self-gravity, and instead confine the gas by adding a fixed gravitational potential, which is constant in $x$ and $y$ and in the $z$ direction has a minimum at the centre of the computational domain, which we denote as $z = 0$. In \citetalias{QEDI} we took this potential to be that of the stars and dark matter only, a reasonable approximation for Solar neighbourhood conditions where the stellar plus dark matter surface density exceeds the gas surface density by a factor of $\approx 5$. However, for our $\Sigma$2.5 conditions, which are intended to represent dwarf- or outer spiral-like environments that are observed to be gas-dominated, it is no longer reasonable to ignore gas gravity. We therefore add a term to our fixed potential to represent this effect.

Our expression for the stellar and dark matter potential is identical to the one used in \citetalias{QEDI}, which we reproduce here for convenience:
\begin{equation}
    \Phi_{\rm *-dm} =
    2\pi G \Sigma_* z_* \left[ \left( 1 + \frac{z^2}{z_*^2} \right)^{1/2}-1\right]
    + 2\pi G \rho_{\rm dm} R_0^2 \rm{ln}\left(1+ \frac{z^2}{R_0^2}\right)\,.
    \label{eq:Phiext}
\end{equation}
Here, $\Sigma_*$, $z_*$, $\rho_{\rm dm}$, $R_0$ are the surface density of stars, scale-height of the stellar disc, the central density of dark matter, and a notional galactocentric radius, which controls how quickly the spherical-shaped dark matter potential falls off with $z$. For the purposes of enabling convenient comparison with SMAUG \citep{Kim+20}, we use the same values of $\Sigma_*$, $z_*$, $\rho_\mathrm{dm}$, and $R_0$ as they choose for runs of equal gas surface density; this is $z_* = 245$ pc in all cases, and we report the remaining values for all runs in \autoref{tab:params}. Note that the gas fraction is 20\% for the $\Sigma$13 and $\Sigma$50 runs, but rises to 60\% in the $\Sigma$2.5 series, as noted above.

We add gravity due to gas, and simultaneously determine the initial vertical profile of the gas density, as follows. The gas in the domain comprises two distinct phases with velocity dispersions $\sigma_1$ and $\sigma_2$, related by $\sigma_2 = 10\sigma_1$, with essentially all of the mass in the first component; as with our other parameters describing the potential, we adopt the same values of $\sigma_1$ and $\sigma_2$ as SMAUG to enable direct comparison (\autoref{tab:params}). We find the density profile and contribution to the gravitational potential from component 1 by solving the equation of hydrostatic equilibrium and the Poisson equation simultaneously,
\begin{equation}
    \sigma_1\frac{d \rho_1}{dz} =
    \rho_1 (g_1 + g_{\rm *-dm}) \qquad
    \frac{d g_1}{dz} = 4 \pi G \rho_1,
    \label{eq:gas_eqns}
\end{equation}
where $\rho_1$ and $g_1$ are the density and gravitational acceleration due to gas component one, and $g_{\rm *-dm} = -d\Phi_{\rm *-dm}/dz$ is the gravitational acceleration due to stars plus dark matter. We solve these equations subject to the constraints that $g_1 (z=0)=0$ and that $\Sigma_{\rm gas} = 2 \int_0^{\infty} \rho_1 dz$. The second constraint requires an iterative approach, so we take an initial guess value for $\rho_1(z=0) \equiv \rho_{1,0}$, solve the equations and integrate to find $\Sigma_{\rm gas}$, and then iteratively adjust the guess until we find the value of $\rho_{1,0}$ that yields the desired value of $\Sigma_{\rm gas}$. We report the resulting value of $\rho_{1,0}$ in \autoref{tab:params}. 

For the second gas component we set the midplane density $\rho_{2,0} = 10^{-5}\rho_{1,0}$, and then determine the profile in $z$ again using a simultaneous solution of the Poisson and hydrostatic equilibrium equations,
\begin{equation}
    \sigma_1\frac{d \rho_2}{dz} =
    \rho_2 (g_2 + g_1 + g_{\rm *-dm}) \qquad
    \frac{d g_2}{dz} = 4 \pi G \rho_2\,,
\end{equation}
where $\rho_2$ and $g_2$ are the density and gravitational acceleration due to gravity from the second component. Since $\rho_{2,0}$ is fixed, no iteration is required. Note that our procedure for the first component is not fully self-consistent, since in principle $g_2$ should appear in \autoref{eq:gas_eqns} as an additional acceleration term inside the parentheses. However, since $g_1 \gg g_2$, ignoring this correction is a reasonable approximation.

Once we have determined the density profiles for each component, we initialise the total gas density profile to the sum of the two, the gas velocity to zero, and the gas pressure as $P= \rho_1 \sigma_1^2 + \rho_2 \sigma_2^2$ which sets the temperature to $\sim 10^4$ K. We set the acceleration of the gravitational field that confines our gas to $g_1 + g_2 + g_{\rm *-dm}$. In principle we could reduce the gravitational potential of the gas over time as outflows deplete the domain, but we show below that this is not a major effect, and since the gas is already subdominant compared to the stars, we omit it for simplicity.

\subsection{Supernovae, metal injection, and cooling}\label{subsec:sn_metal_cooling}

We set off SN explosions in our computational domain using the same procedure as in \citetalias{QEDI}. We derive a star-formation surface density $\dot{\Sigma}_*$ from the Kennicutt-Schmidt law \citep{Kennicutt98a} using the $\Sigma_{\rm gas}$ value of each run (reported in \autoref{tab:params}) and then estimate a SN surface density from this assuming a \citet{Chabrier2001} initial mass function, which gives a SN rate in the computational domain $\Gamma_\mathrm{SN} = \dot{\Sigma}_* A / M_\mathrm{per-SN}$, where $M_\mathrm{per-SN} = 100$ M$_\odot$ is the stellar mass require to produce one SN and $A = 1$ kpc$^2$ is the area of the simulation domain. The SNe are distributed uniformly in the $x-y$ plane of the galaxy while their $z-$distribution is Gaussian with a scale height $h_\mathrm{SN}$ that varies by run. We implement SNe using the same procedure as described in \citetalias{QEDI}: for each time step of size $dt$, we determine a number of SNe that will occur by drawing from a Poisson distribution with expectation value $\Gamma_\mathrm{SN} \, dt$, and then choose a location for each SN by drawing from uniform distributions in the $x$ and $y$ directions and from a Gaussian of the width $h_\mathrm{SN}$ in the $z$ direction. For every SN that goes off, we add $\Delta E_\mathrm{SN} = 10^{51}$ erg of thermal energy to the cell that contains its position; we also add to that cell a total mass $\Delta M_\mathrm{SN} = 5$ \msun\ and a metal mass $\Delta M_{Z,\mathrm{SN}} = 1$ M$_\odot$ to the cell to represent the total and metal mass in SN ejecta\footnote{Note that the total amount of mass added to the domain by this process is negligible. Quantitatively, the fractional increase in the gas mass in the domain added by SNe is $(\Sigma_\mathrm{SFR} t_f/\Sigma_\mathrm{gas})(\Delta M_\mathrm{SN}/M_\mathrm{per-SN})$, and this value is $<1\%$ for all runs.}; the metals are treated as a passive scalar, and our choice of $\Delta M_{Z,\mathrm{SN}}$ is a typical yield of oxygen from a type II SN. The passive tracer then evolves along with the rest of the simulation. Note that our treatment here differs slightly from the procedure in \citetalias{QEDI}, where we added only energy and metals, but no mass. Whether we include the mass of the SN ejecta or not has little effect on the outcome of the simulations, but we find that including it improves code performance by avoiding the occasional production of cells with very high temperatures that necessitate small time steps, as well as being more realistic.

The heating of the gas provided by SNe is offset by radiative cooling, which we include as a customized source term in the energy equation that is similar, but not identical to, \textsc{Grackle} library \citep{Grackle}. We direct the interested reader to the Methods section of \qedi\ for a details on the implementation. Our cooling function includes primordial, metal line, and Compton cooling together with photoelectric heating. We calculate the latter assuming a uniform \cite{Haardt_2012} UV background. Our treatment here is mostly identical to that in \citetalias{QEDI}, and we direct the readers to Section 2.1 of that paper for more details. The one place where our approach here differs is that \citetalias{QEDI} adopted metal line  cooling and photoelectric heating rates computed for Solar metallicity, while in this paper we multiply those rates by a factor $Z_{\rm bg}/Z_{\odot}$, where $Z_\mathrm{bg}$ is the metallicity we adopt for each run -- $Z_\odot$ for the Z1 runs, and $0.2Z_\odot$ for the Z0.2 runs; primordial and Compton cooling are left unchanged. Note that this is not fully self-consistent, in the sense that the cooling rate does not reflect increases in the local gas metallicity due to SN ejecta, only the mean metallicity at the start of the simulations. While this approach is not fully realistic, it has the advantage that it allows us to perform clean experiments where we can hold the metallicity that goes into cooling fixed; if we did not do this, then the low-metallicity runs would relatively quickly converge with the higher-metallicity ones due to self-enrichment, muddying efforts to isolate the effects of metallicity. We explore the implications of our treatment of cooling in \aref{app:cooling}.

\subsection{Analysis: outflow rates and loading factors}\label{subsec:outflow_rates}

The primary quantities we wish to extract from the simulations are the outflow rates of mass, metals, and energy as a function of height $z$. We define these, respectively, as
\begin{equation}\label{eqn:mdot}
    \left(
    \begin{array}{c}
    \dot{M} \\
    \dot{M}_Z \\
    \dot{E}
    \end{array}\right)
    = \int_0^L \int_0^L 
    \left[
    \left(\begin{array}{c}
    \rho v_z \\
    \rho_Z v_z \\
    \rho e v_z
    \end{array}
    \right)_{+z} - 
    \left(\begin{array}{c}
    \rho v_z \\
    \rho_Z v_z \\
    \rho e v_z
    \end{array}
    \right)_{-z}
    \right]\,dx \, dy\,,
\end{equation}
where $\rho$, $\rho_Z$, $\rho e$, and $v_z$ are the mass density, metal density, total (thermal plus kinetic) energy density, and $z$ component of gas velocity, and the computational domain extends from 0 to $L=1$ kpc in the $x$ and $y$ directions. We note that these definitions include contributions from both outflowing and inflowing gas, not only from gas with outward velocities. Also note that our definition sums over the $+z$ and $-z$ sides of the galactic disc, as indicated by the $+z$ and $-z$ subscripts for the quantities in parentheses.

It is convenient to normalise these outflow rates to the rates to rates of star formation, metal injection, and energy injection, respectively. We define the loading factors by
\begin{equation}
    \left(\begin{array}{c}
    \eta_M \\
    \eta_Z \\
    \eta_E
    \end{array}
    \right) = \Gamma_\mathrm{SN}^{-1} \left(
    \begin{array}{c}
    \dot{M} / M_\mathrm{per-SN} \\
    \dot{M}_Z / \Delta M_{Z,\mathrm{SN}} \\
    \dot{E} / \Delta E_\mathrm{SN}
    \end{array}
    \right).
    \label{eqn:mass_loading}
\end{equation}
We remind readers that each of these quantities is a function of height $z$ and time $t$. 

The metal loading factor necessitates a bit more discussion. The quantity we have defined is the instantaneous loading factor that is often measured in simulations, but this is not the quantity of greatest interest for the purposes of galactic chemical evolution models. As discussed in \citetalias{QEDI}, the quantity of interest for this purpose is the fraction of newly-injected metals that are promptly lost to the wind rather than being retained in the ISM, $\phi$ \citep[e.g.,][]{Sharda21a, Sharda21b}. We measure \fy\ from our simulations using $\dot{M}$ and $\dot{M}_Z$,  the fluxes of total and metal mass through a given surface of height $z$ (\autoref{eqn:mdot}), and 
\begin{equation}
    \langle Z \rangle = \frac{\int_{-z}^z \int_0^L \int_0^L \rho_Z \, dx \, dy \, dz}{\int_{-z}^z \int_0^L \int_0^L \rho \, dx \, dy \, dz}\,,
\end{equation}
which is the mean metallicity of the gas closer to the midplane than $z$. Given this definition, we now define
\begin{equation}
    \phi = \frac{\dot{M}_Z - \langle Z\rangle \dot{M}}{\Gamma_\mathrm{SN} \,\Delta M_{Z,\mathrm{SN}}} = \left(\frac{\zeta-1}{\zeta}\right) \eta_Z,
    \label{eq:phi}
\end{equation}
where $\zeta=\dot{M}_Z/\langle Z \rangle \dot{M}$ is the ratio of the wind metal flux to the metal flux that would be expected if the wind consisted purely of ISM at the mean metallicity of the gas. Thus intuitively we can understand the numerator in this expression as the difference between the actual metal outflow rate and the rate that would be expected for an outflow with the same metallicity as the ISM from which it is generated, i.e., it is the \textit{excess} metal content in the outflow compared to the case of a fully-mixed outflow. In turn, this means that we can interpret $\phi$ as a modified version of the simple metal loading factor $\eta_Z$ that is corrected by removing the contribution to metal loading from entrained ISM, leaving only the direct SN contribution. 

\section{Results}\label{sec:results}

We will first discuss results from the $\Sigma$2.5-Z1-H150, $\Sigma$13-Z1-H150, and $\Sigma$50-Z1-H150 runs, where we keep the metallicity and SN scale height constant and vary the gas surface density, in \autoref{subsec:var_sigma}. We then compare to the Z0.2 runs to study the effect of varying galaxy metallicity (\autoref{subsec:var_Z}) and the $\Sigma$2.5-Z1-H* runs to study the effect of varying the height of SN injection (\autoref{subsec:var_H}).

\subsection{Surface density variation}
\label{subsec:var_sigma}

\subsubsection{Qualitative outcomes}

\begin{figure*}
    	\includegraphics[width=0.65\textwidth]{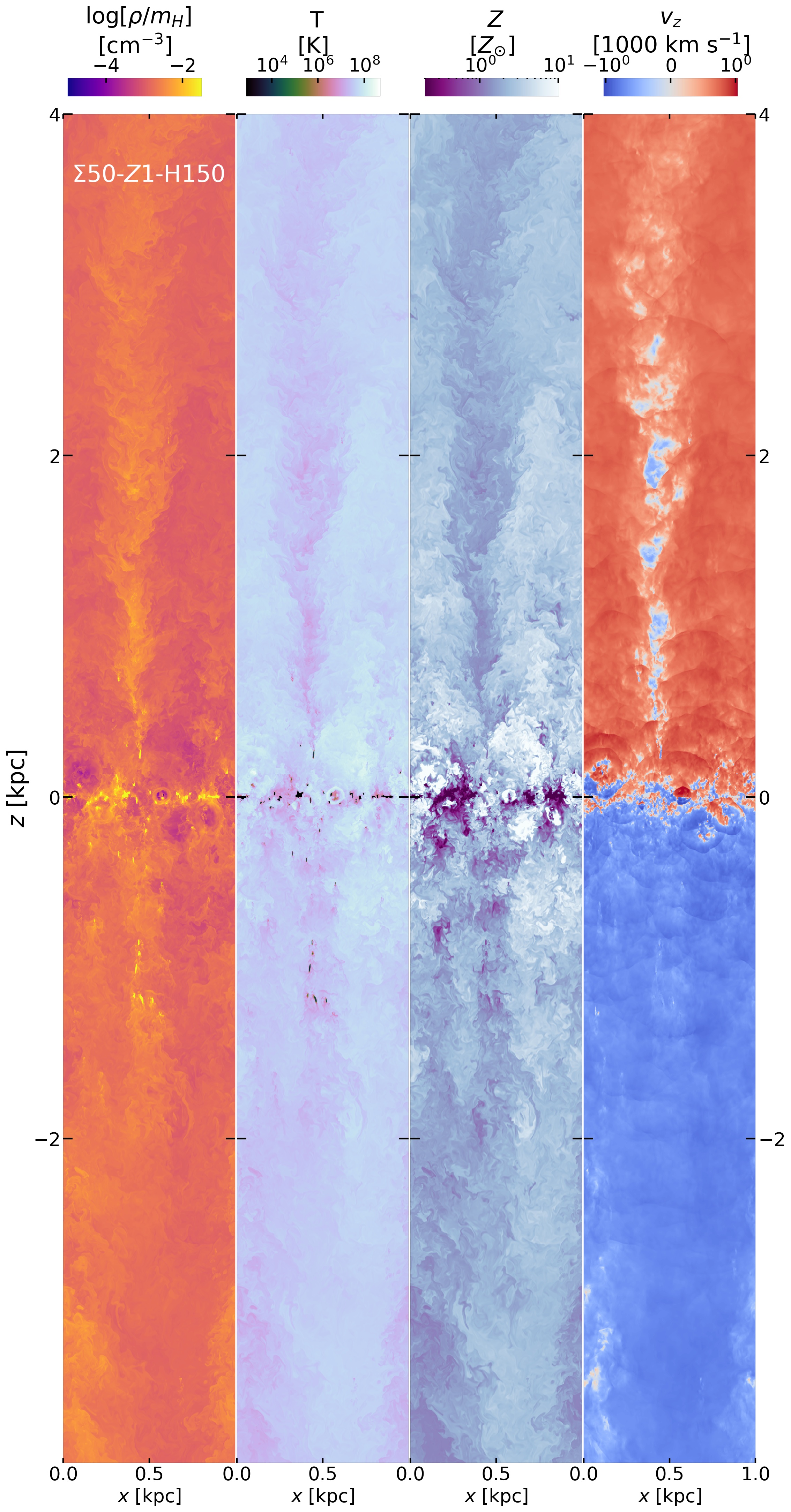}
    \caption{A slice through run $\Sigma$50-Z1-H150. From left to right, the quantities shown are gas density, temperature, injected metal abundance normalised to Solar, and $z$ velocity. }
    \label{fig:slice_sig4}
\end{figure*}

\begin{figure*}
    	\includegraphics[width=0.7\textwidth]{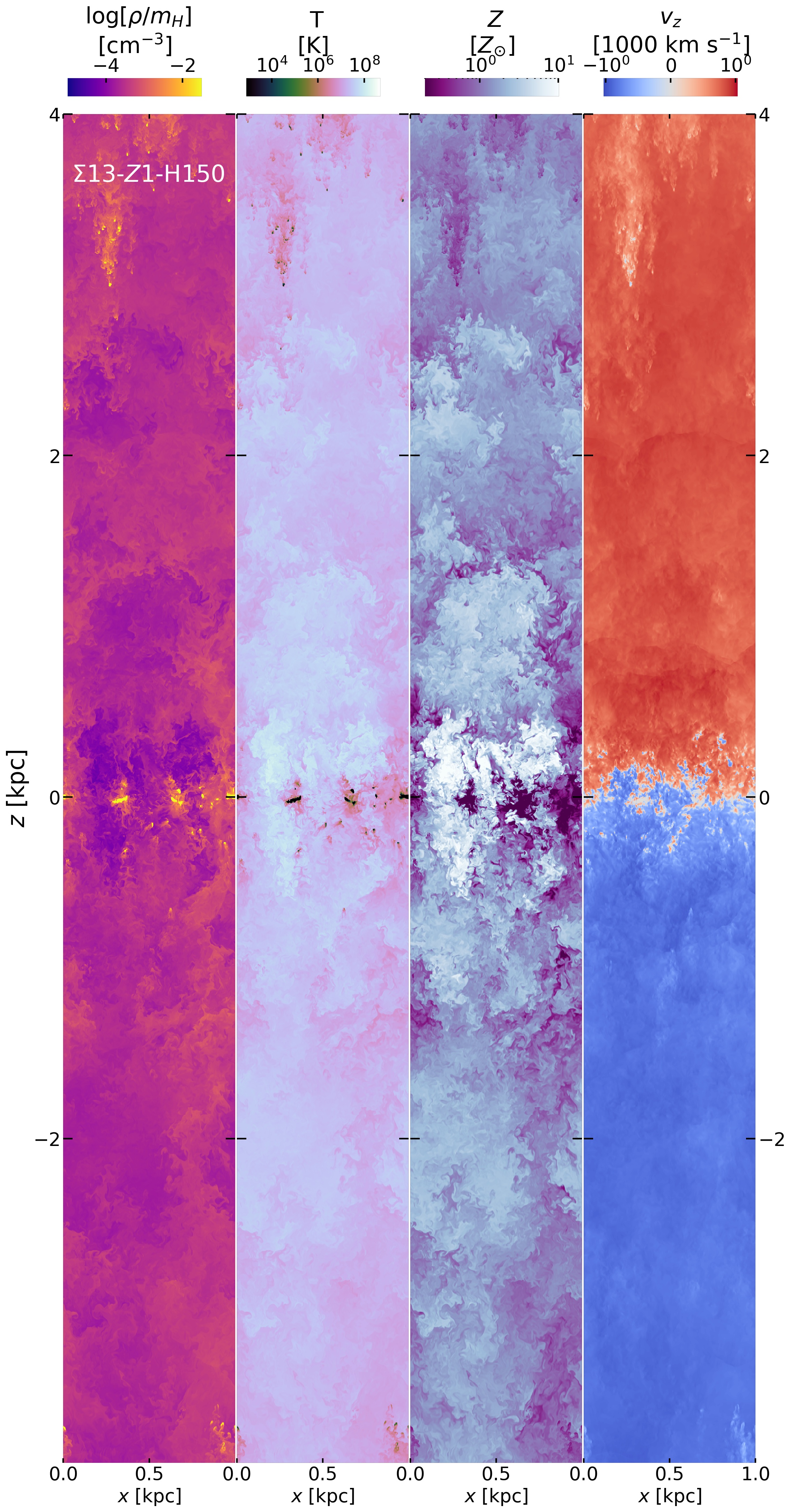}
    \caption{Same as \autoref{fig:slice_sig4}, but for the $\Sigma$13-Z1-H150 run.}
    \label{fig:slice_sig}

\end{figure*}

\begin{figure}
    	\centerline{\includegraphics[width=0.75\columnwidth]{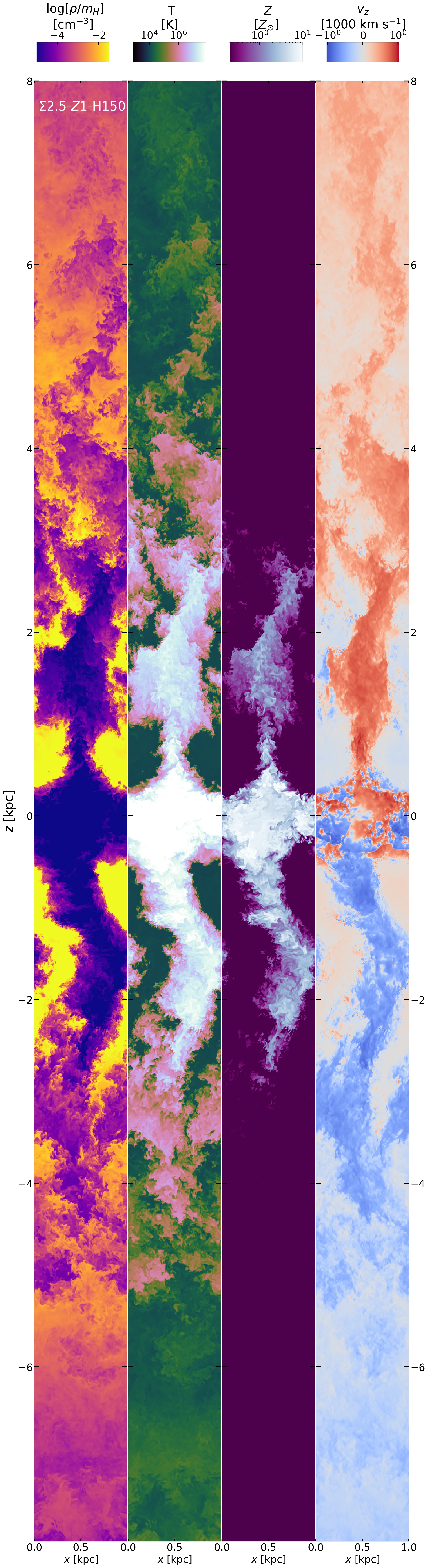}}
    \caption{Same as \autoref{fig:slice_sig4}, but for the $\Sigma$2.5-Z1-H150 run; note that this run uses a box half-height $L_z$ twice as large as the simulations shown in \autoref{fig:slice_sig4} and \autoref{fig:slice_sig}.}
    \label{fig:slice_0.2sig}

\end{figure}

In \autoref{fig:slice_sig4}, \autoref{fig:slice_sig} and \autoref{fig:slice_0.2sig} we show $y=0$ slices from runs $\Sigma$50-Z1-H150, $\Sigma$13-Z1-H150, and $\Sigma$2.5-Z1-H150, respectively, at times after the simulation has settled into steady state (see below). From left to right in all the plots we show density, temperature, injected metal abundance scaled to Solar, and $v_z$. 

From the outset, we note that the outflow structure is very different for the three gas surface densities. For $\Sigma$50-Z1-H150, the higher gas surface density and consequently the larger star formation rate ensures that the outflows are hot and devoid of warm clumps. 
The mean metallicity of the volume-filling hot gas is highest near the midplane, where it consists predominantly of almost pure SN ejecta, and falls with height as the hot phase is diluted by cooler, lower-metallicity entrained gas mixing into it. Most of the gas is outflowing at $\sim 100$ \kmps\ and once steady state is reached the gas is outflowing for the duration of the run.

Run $\Sigma$13-Z1-H150 is the Solar neighbourhood case, and produces a multiphase outflow with a very wide range of gas temperatures and densities. There is good phase separation as there are sharp vertical gradients in density, temperature, metallicity and velocity between the cool-to-warm clumps and the hot volume-filling medium. 
As in the $\Sigma$50-Z1-H150 case, we see a clear decrease in the mean metallicity of the hot volume-filling phase as we move away from the plane, but in this run there is also a much more pronounced gradient in temperature and velocity, with the outflow hottest and fastest closest to the plane and cooling and slowing as it moves upward. Unlike in the $\Sigma$50-Z1-H150 run, in this case we see clear clumps of cooler gas embedded in the hot flow far from the plane; some of these are moving outward, albeit more slowly than the hot phase, but some have begun to fall back toward the plane in a fountain flow. This fallback process is highly stochastic and therefore the outflow rates have high temporal variability. 

\autoref{fig:slice_0.2sig} shows that $\Sigma$2.5-Z1-H150 case has structure very different from the other two. In this case, hot gas is continuously produced near the midplane but the production rate is too low from the hot gas to punch channels through the warm gas as happens in the other two runs. This leads to a positive feedback loop in that the failure of the SNe to produce a volume-filling hot phase means that most of the SNe explode in a relatively dense environment, which further impedes the ability of the hot gas to expand and become volume-filling. As a result SNe are not able to launch substantial outflows, and this ensures metals remain trapped close to the midplane. The hot phase is not able to transfer significant momentum to the ambient gas which has a lot of inertia and therefore is slow. 

\subsubsection{Outflow rates and loading factors}

\begin{figure*}
    	\includegraphics[width=\textwidth]{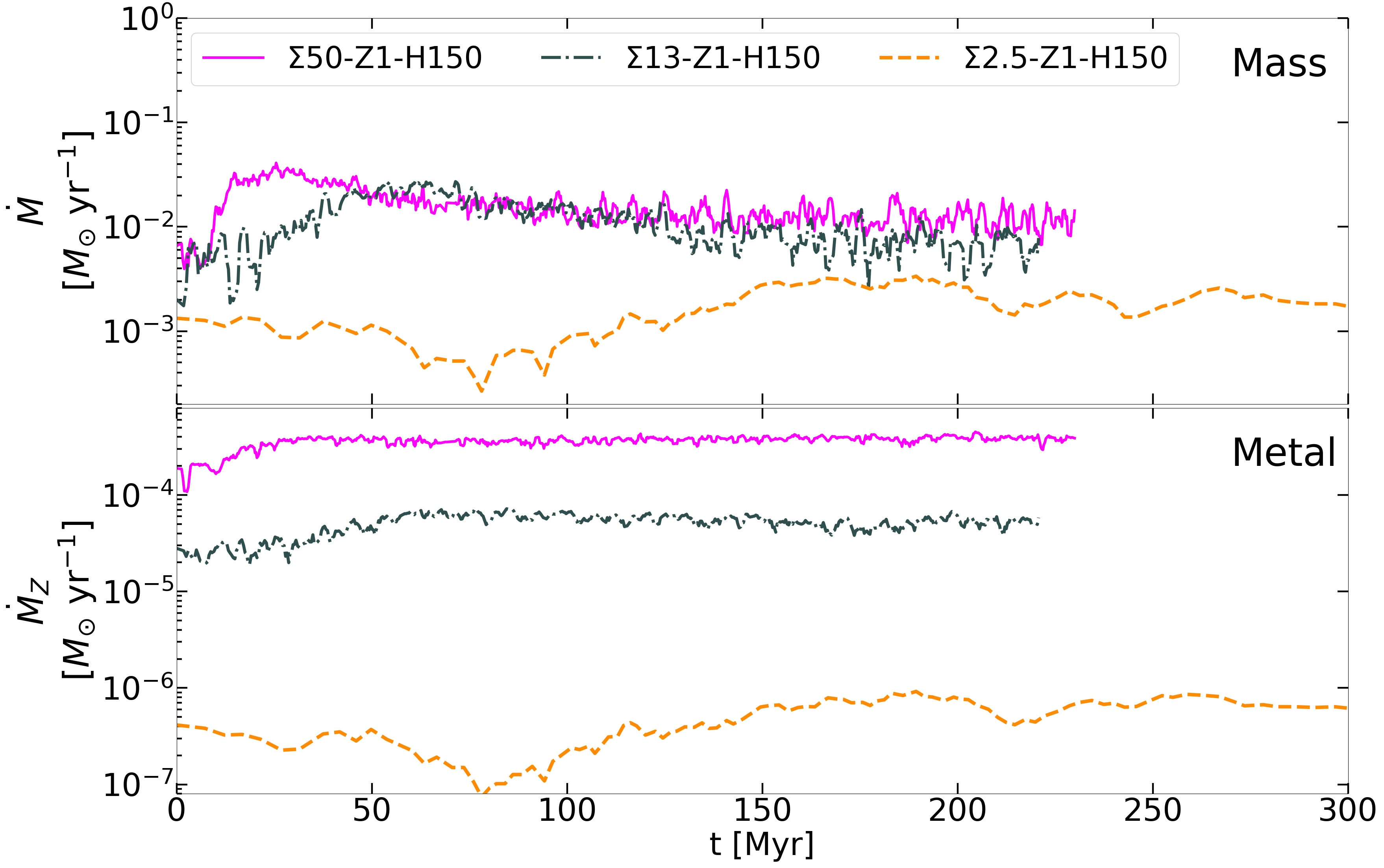}
    \caption{Mass (top) and metal (bottom) outflow rates for runs $\Sigma$50-Z1-H150, $\Sigma$13-Z1-H150, and $\Sigma$2.5-Z1-H150 at a distance of $L_z/2$ from the midplane. The outflow rate is integrated over the full simulation domain in $x$ and $y$, and is the sum of the values at $\pm L_z/2$ -- see \autoref{eqn:mdot}. The outflow rates reach a steady value well before the end of the simulation. The mass outflow rates for $\Sigma$50-Z1-H150 and $\Sigma$12.5-Z1-H150 are comparable while the metal outflow rate is much higher for $\Sigma$50-Z1-H150.  
    }
    \label{fig:outflow_rates}

\end{figure*}

\autoref{fig:outflow_rates} shows the mass (top) and metal (bottom) outflow rates (see \autoref{eqn:mdot}) at distance of $L_z/2$ from the midplane as a function of time. The plots indicate that the outflow properties of the system have reached a steady-state over the duration of the simulations, at least for $\Sigma$50-Z1-H150 and $\Sigma$13-Z1-H150, where there is a significant outflow at all times. As one might have expected based on the slice plots, the metal outflow rate for $\Sigma$50-Z1-H150 is higher than that for $\Sigma$13-Z1-H150, while their mass outflow rates are comparable, and that the $\Sigma$13-Z1-H150 outflow rates are comparable to the ones reported in \qedi, despite the small changes we have made in simulation procedure.

\begin{figure*}   	\includegraphics[width=\textwidth]{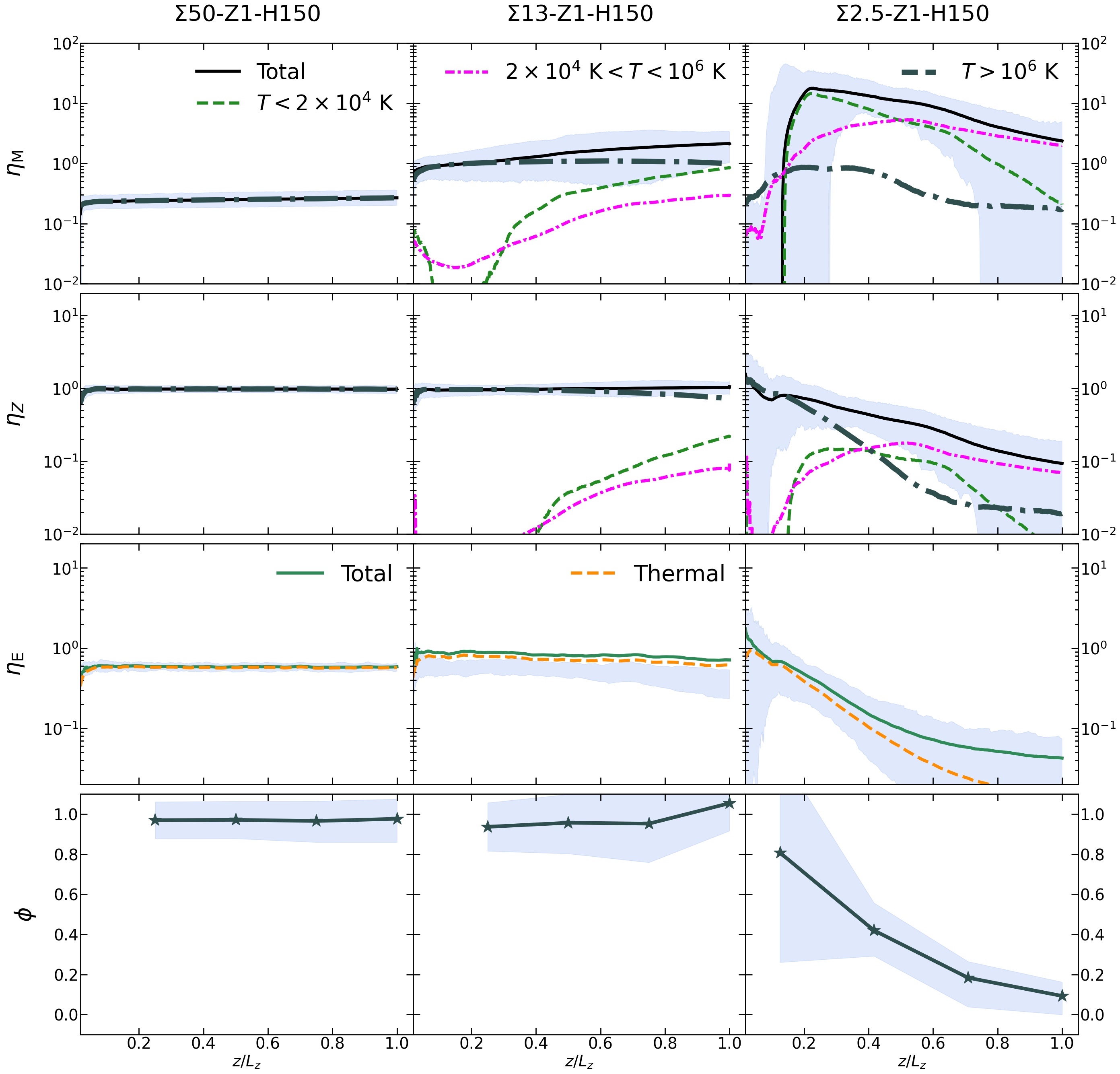}
    \caption{Mass, metal, and energy loading factors ($\eta_M$, $\eta_Z$, $\eta_E$, top three rows) and fraction of injected metals lost promptly to the wind ($\phi$, bottom row) as a function of height $z$ normalised to box half-height $L_z$ for the $\Sigma$50-Z1-H150, $\Sigma$13-Z1-H150, and $\Sigma$2.5-Z1-H150 runs (left to right columns). Solid curves show time averages for $t>75$ Myr while blue bands indicate the $16$th to $84$th percentile variation over time. The dotted and dashed lines in the top two rows for $\eta_M$ and $\eta_Z$ show the contributions from the cool ($T<10^4$ K, green dashed), warm unstable ($10^4\;\mathrm{K} < T < 10^6\;\mathrm{K}$, magenta dot-dashed) and hot  ($T>10^6$ K, black long dot-dashed) phases. In the row for $\eta_E$, we show contribution from the thermal energy density as the orange dashed line. For $\phi$, we show the results computed at four heights $z$.}
    \label{fig:loading_factors}
\end{figure*}

\begin{table*}
\begin{center}
\begin{tabular}{|l|c|c|c|c|c|c|c|c|}
\hline
\hline
Name & $\eta_M$ & $b$ & $\eta_M^{\rm cool}$ & $\eta_M^{\rm warm}$ & $\eta_M^{\rm hot}$ & $\eta_E $& $\eta_Z$ & \fy\\
 
(1)  & (2) & (3) & (4) & (5) & (6) & (7) & (8) & (9)\\
\hline
% \multicolumn{9}{c}{Fixed grid runs} \\
\hline
$\Sigma$13-Z1-H150 & $2.1$ & $1.0$ & $0.82$ & $0.31$ & $1.0$ & $0.64$ & $1.0$ & $0.97$  \\
\\
$\Sigma$50-Z1-H150 &  $0.26$ & $0.58$ & $0.0$ & $0.0$ & $0.26$ & $0.97$ & $0.86$ & $0.96$ \\
\\
$\Sigma$2.5-Z1-H150  &  $2.3$ & $2.0$ & $0.20$ & $1.69$ & $0.17$ & $0.014$ &  $0.093$ & $0.052$  \\

\\
%%Sub-Solar Metallicity
$\Sigma$13-Z0.2-H150 & $0.040$ & $2.7$ & $1.0\times 10^{-5}$ & $0.040$ & $1.4\times 10^{-4}$ & $1.3\times 10^{-3}$ & $6.0\times 10^{-3}$ & $2.7\times 10^{-3}$ \\
\\

$\Sigma$50-Z0.2-H150 & $0.48$ & $0.65$ & $0.0$ & $0.0$ & $0.48$ & $0.48$ & $0.92$ & $0.89$  \\
\\
$\Sigma$2.5-Z0.2-H150 & $1.6$ & $2.1$ & $0.081$ & $1.5$ & $5.5\times 10^{-3}$ & $0.012$ & $0.034$ & $4.5\times 10^{-3}$ \\

\\
%%hSN variation
$\Sigma$2.5-Z1-H300 & $0.28$ & $2.8$ & $8.2\times 10^{-4}$ & $0.27$ & $7.9\times 10^{-3}$ & $8.9\times 10^{-4}$ & $0.017$ & $0.014$ \\
\\
$\Sigma$2.5-Z1-H1000 & $3.2\times 10^{-3}$ & $3.0$ & $-2.7 \times 10^{-5}$  & $3.3\times 10^{-3}$ & $0.0$ & $1.8\times 10^{-5}$ & $5.7\times 10^{-4}$ & $5.0\times 10^{-4}$ \\
\\

$\Sigma$2.5-Z1-H1500 & $4.6$ & $2.3$ & $2.0$ & $2.5$ & $0.044$ & $2.9\times 10^{-3}$ & $0.27$ & $0.18$ \\
\\
$\Sigma$2.5-Z1-H2000 &  $3.6$ & $0.91$ & $0.019$ & $1.7$ & $1.8$ & $0.39$ & $0.62$ & $0.56$ \\

\hline
\hline
\\

\end{tabular}
\caption{Summary of results for all runs. All quantities reported here are averaged for $t>75$ Myr and $z=L_z$. Column 2- the mass loading factor ($\eta_M$). Column 3- the burstiness parameter $b$ (\autoref{eq:bursty}). Columns 4, 5, and 6- the contribution to $\eta_M$ from cool ($T<2 \times 10^4$ K), warm unstable ($2 \times 10^4$ K $<T<10^6$ K), and hot ($T>10^6$ K) gas, respectively. Columns 7 and 8- the energy loading ($\eta_E$) and the metal loading ($\eta_Z$) factors. Column 9- the average fraction of metals promptly lost to the wind rather than retained in the ISM $\phi$ (\autoref{eq:phi}). 
}
\label{tab:results}
\end{center}
\end{table*}

From the outflow rates we compute the loading factors $\eta_M$, $\eta_Z$ and $\eta_E$, along with the fraction of metals lost to the outflow $\phi$, following the procedure described in \autoref{subsec:outflow_rates}. We do this as a function of time for each simulation snapshot at $t>75$ Myr, roughly the point at which the simulations settle into steady state. We then compute the average and 16th to 84th percentile range over time. We use the latter to define a ``burstiness parameter''
\begin{equation}
    b = \frac{\eta_M^{84} - \eta_M^{16}}{\langle\eta_M\rangle}
    \label{eq:bursty},
\end{equation}
where $\eta_M^{84}$ and $\eta_M^{16}$ are the 84th and 16th percentile values and $\langle \eta_M\rangle$ is the time average. This parameter characterises the relative width of the distribution of mass loading factors. We report-time averaged values of each quantity at $z=L_z$ (i.e., at the edge of the simulation box) in \autoref{tab:results}, and plot these values as a function of $z$ in \autoref{fig:loading_factors}.\footnote{A further implication of the values of $\eta_M$ reported in \autoref{tab:results} is that none of our runs lose a large fraction of their initial gas mass to outflows. The fraction of the mass lost to that initially present in the domain is $\eta_M \Sigma_\mathrm{SFR} t_f/\Sigma_\mathrm{gas}$; this fraction is 22\% for run $\Sigma50$-Z1-H150, 11\% for $\Sigma$50-Z0.2-H150, and below 10\% for all other runs.} In this figure solid lines represent averages and shaded bands showing the $16$th to $84$th percentile range; we further subdivide $\eta_M$ and $\eta_Z$ into contributions to the mass and metal flux from cool ($T<2 \times 10^4$ K, green dashed), warm unstable ($2 \times 10^4\;\mathrm{K} < T < 10^6\; \mathrm{K}$, pink dot-dashed), and hot ($T>10^6$ K, grey thick dot-dash) phases, and $\eta_E$ with total (green) and thermal (orange dashed) contributions to the energy flux separately.

The variations in $\eta_M$ and $\eta_E$ across the three runs reflect the varying outflow structure visible in the slice plots. For the highest surface density case, $\Sigma$50-Z1-H150, we find weak mass loading, $\eta_M \approx 0.26$, with little variation with time or height ($b \approx 0.6$), but efficient energy loading, $\eta_E \approx 0.97$. Mass-loading is dominated by the hot phase, and energy loading by the thermal component. This is not surprising since the outflows (\autoref{fig:slice_sig4}) possess very few warm clouds compared to $\Sigma$13-Z1-H150, our Solar neighbourhood analog. The $\Sigma$13-Z1-H150 case shows higher overall mass loading ($\eta_M \approx 2.1$) and lower energy loading ($\eta_E = 0.64)$, and larger temporal variation in mass-loading ($b\approx 1$). This burstiness is also reflected in the energy loading, which is dominated by rare events so that the mean over time is above the 16th to 84th percentile range. Cool and hot gas make roughly equal contributions to the mass flux, consistent with the multiphase structure visible in \autoref{fig:slice_sig}; the contribution to the outflow rate from cooler gas increases with height, but as shown in the analysis presented in \citetalias{QEDI}, this does not reflect cooling of hot gas; instead, it reflects acceleration of pre-existing cool clouds by the hot gas, which leads to an increasing velocity and thus an increasing contribution to mass flux with height. Finally, $\Sigma$2.5-Z1-H150 is even more heavily mass-loaded and bursty ($b\approx 2)$, with the mass loading factor at its largest close to the disc, but remaining larger than for $\Sigma$13-Z1-H150 even at 8 kpc, $\eta_M \approx 2.3$. The mass flux is dominated by cool gas out to $\approx 4$ kpc, transitioning to warm unstable gas at larger heights, and with no significant contribution from hot gas at any height. Due to the dearth of hot gas the energy loading is small, $\eta_E \approx 0.014$, and dominated by kinetic energy except near the midplane, and outflows are rather slow, as is clear from examining \autoref{fig:slice_0.2sig}.

The metal loading factor $\eta_Z$ and fraction of metals lost to outflow $\phi$ are high and nearly-independent of height in $\Sigma$50-Z1-H150 and $\Sigma$13-Z1-H150. Our finding that most metals are lost to the wind is in line with the results of \qedi. As with $\eta_M$, there is little time variation in $\phi$ for $\Sigma$50-Z1-H150, and relatively more for $\Sigma$13-Z1-H150. By contrast, for $\Sigma$2.5-Z1-H150 most of the injected metals remain within the box, and $\phi\lesssim 0.1$ even at large heights; what metals are lost are primarily carried by the warm unstable gas phase. The gas undergoes cycles of puffing up and subsequent relaxation but few metals are able to escape, particularly in comparison to the relatively large mass loading factor for this run.

\subsection{Metallicity variation}
\label{subsec:var_Z}

\begin{figure*}
    	\includegraphics[width=\textwidth]{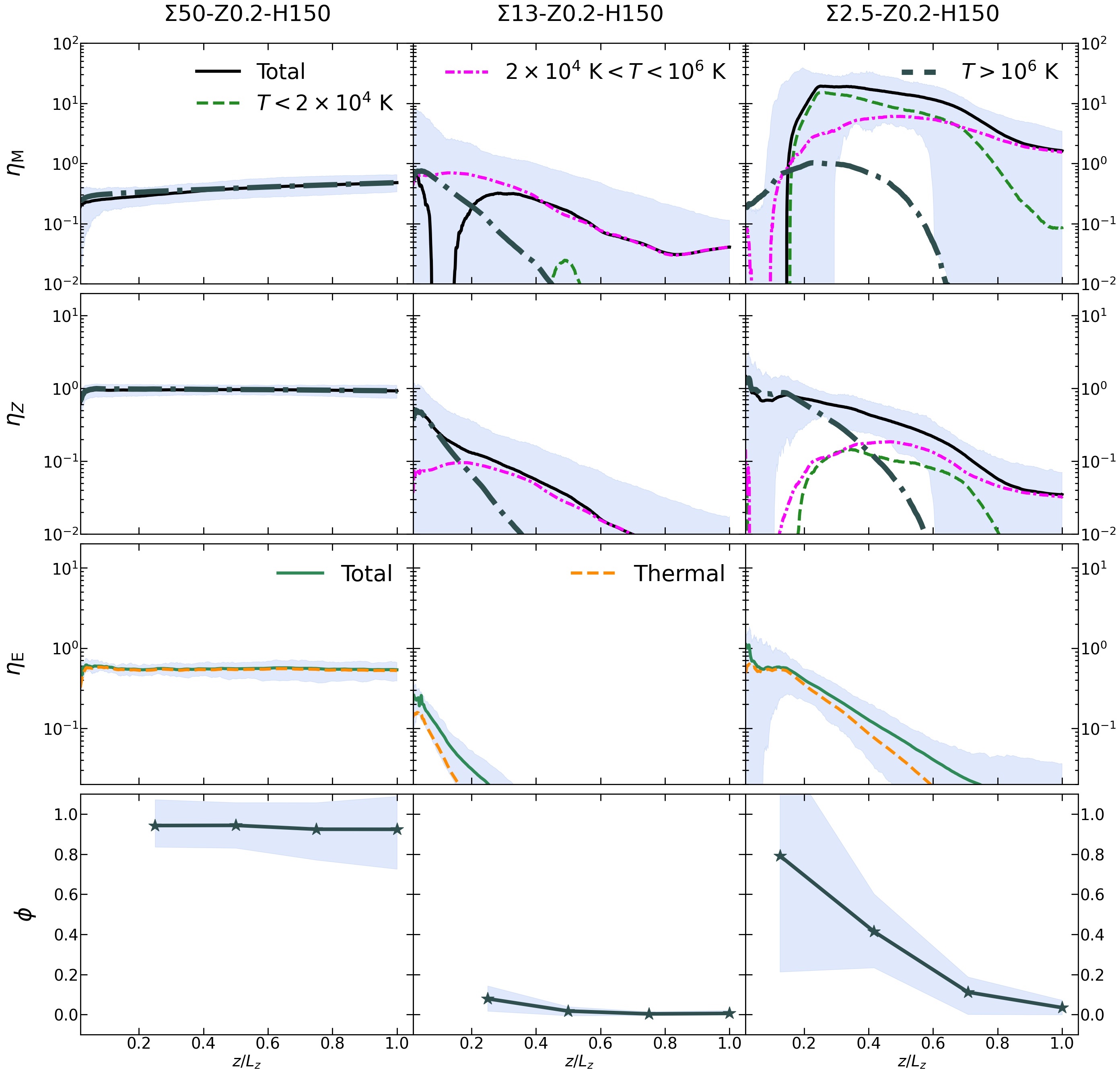}
    \caption{Same as \autoref{fig:loading_factors} except now we show $\Sigma$XX-Z0.2-H150 series.}
    \label{fig:loading_factors_0.2Zsol}

\end{figure*}

In the runs described thus far -- $\Sigma$50-Z1-H150, $\Sigma$13-Z1-H150, and $\Sigma$2.5-Z1-H150 -- our cooling function was set to a value appropriate for Solar abundances. We now compare those results to the results of runs $\Sigma$50-Z0.2-H150, $\Sigma$13-Z0.2-H150, and $\Sigma$2.5-Z0.2-H150, which use identical setups in all respects except that the cooling function has been set to that for gas with 20\% Solar metallicity.

\autoref{fig:loading_factors_0.2Zsol} is constructed in exactly the same way as \autoref{fig:loading_factors}, and shows the loading factors and fraction of metals lost promptly to the wind, in the runs with $Z=0.2 Z_{\odot}$. Comparing the two figures, we see that changing the cooling rate has little effect on the $\Sigma$50 or $\Sigma$2.5 runs -- the outcomes for $\Sigma$50-Z1-H150 and $\Sigma$50-Z0.2-H150 are qualitatively very similar, as are the $\Sigma$2.5-Z1-H150 and $\Sigma$2.5-Z0.2-H150 cases. By contrast the $\Sigma$13 runs at $Z=Z_\odot$ and $Z=0.2Z_\odot$ are quite different. Run $\Sigma$13-Z1-H150 has loading factors that are nearly constant with height and only moderately variable with time ($b\approx 1$), with a hot component providing a substantial fraction of the mass flux and most of the metal flux, high energy loading dominated by the thermal component, and most of the metals lost promptly to the wind. By contrast $\Sigma$13-Z0.2-H150 has loadings that decrease strongly with height, with most of the mass and metal flux in the unstable warm phase, a great deal of time-variability ($b\approx 2.7$), low energy loading, and little metal loss.

The difference is due to the cooling behaviour of the ambient cool gas into which the SNe explode. In the $Z=Z_\odot$ case, the cool gas at $T\lesssim 10^4$ K cools very rapidly, leading it to break up into cold clumps, $T \sim 100$ K, that rapidly fall toward the midplane and leave a very low-density volume-filling medium in between.
%\bennotes{This suggests that entropy and buoyancy are critical ingredients in understanding galactic winds! We should explicitly state this (maybe even in the abstract) and also cite work by Mark Voit, Ben Keller, and other people on this subject.} 
By contrast in the runs with $Z=0.2Z_\odot$ the gas cools much more slowly, and instead forms a volume-filling, pressure-supported medium with $T\sim 0.5-1\times 10^4$ K. Quantitatively, we find that in $\Sigma$13-Z1-H150, in the region $|z|<300$ pc, cold gas with $T<500$ K and warm neutral and ionised gas with $500 < T < 2\times 10^4$ K occupy roughly equal volumes (omitting the large fraction also filled by much hotter gas), while in $\Sigma$13-Z0.2-H150 cold gas with $T<500$ K occupies $<1\%$ of the volume occupied by the warm neutral and ionised phases. This difference means the typical density of the medium into which SNe explode in $\Sigma$13-Z0.2-H150 is denser than in $\Sigma$13-Z1-H150. Alternately, one could express this difference as one of entropy, consistent with earlier work finding that the difference in entropy between SN-heated bubbles and their environments is a key driver of galactic winds \citep{Keller20a}. The more rapid cooling in the Z1 run guarantees that most SNe explode in a low-entropy environment, while in Z0.2 the environment is much higher entropy. Regardless of whether one views density or entropy as the key controlling parameter, the effect is that for the Z0.2 case the SNe have trouble breaking out and generating a fast, hot wind that can accelerate cool clouds efficiently at large heights, leading to loading factors that decrease strongly with height rather than remaining flat as in the $Z=Z_\odot$ case, and to what outflow there is being dominated by intermediate temperature gas rather than a distinct mix of hot and cool. The lack of an efficiently-driven hot wind that does not mix with the cooler ambient ISM is also what is responsible for the small fraction of metal loss, since it is this hot, unmixed wind that is responsible for producing efficient metal loading.

Of course the same changes in cooling between $Z=Z_\odot$ and $Z=0.2Z_\odot$ occur in the $\Sigma$50 and $\Sigma$2.5 cases, but here they make much less difference. In the case of $\Sigma$50 this is because, even though the ambient medium cools less efficiently for $Z=0.2Z_\odot$, the stronger gravitational potential is nonetheless able to drag the ambient gas toward the midplane rapidly, ensuring the SNe continue to explode in a low-density medium where they can build up an efficient hot wind. Conversely, in the $\Sigma2.5$ the weaker gravitational potential ensures that even with efficient cooling, $Z=Z_\odot$, the ambient gas is not efficiently pulled to the midplane, so that even in $Z=Z_\odot$ case SNe often encounter dense regions and cannot driven an efficient wind.

\subsection{Variations in  SN scale height}
\label{subsec:var_H}

\begin{figure}
    	\includegraphics[width=\columnwidth]{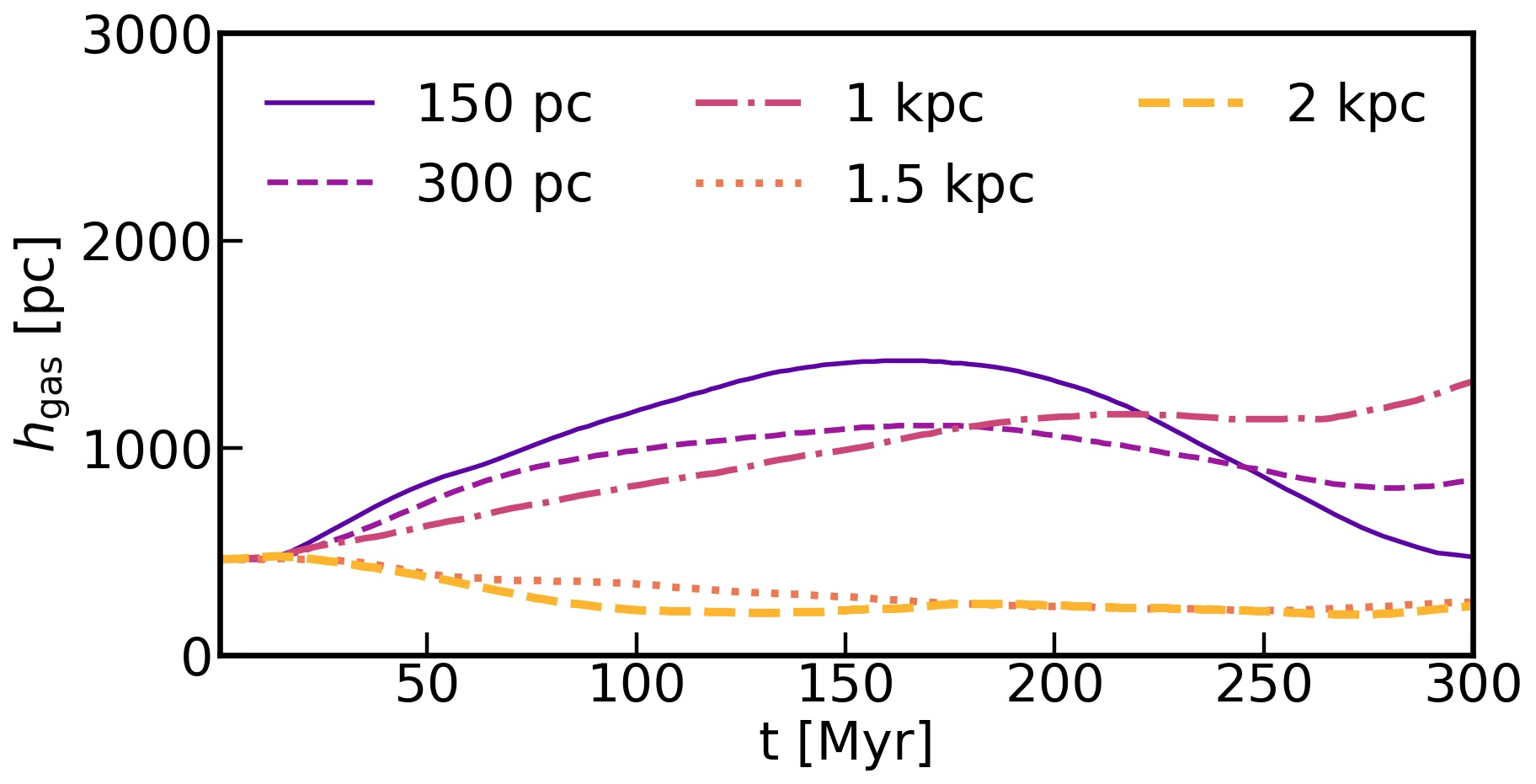}
    \caption{Time averaged gas scale height for $\Sigma2.5$-$Z1$-Hhhh runs where hhh$= 150$ pc, $300$ pc, $1$ kpc, $1.5$ kpc, and $2$ kpc is the SN scale height indicated in the legend. Gas scale height is determined by estimating the height which enclosed 1/e times total gas mass in the box. The different curves represent gas scale heights achieved in the simulation if the SN went off with a scale height indicated. }
    \label{fig:gas_scale_height}

\end{figure}

\begin{figure*}
    	\includegraphics[width=0.7\textwidth]{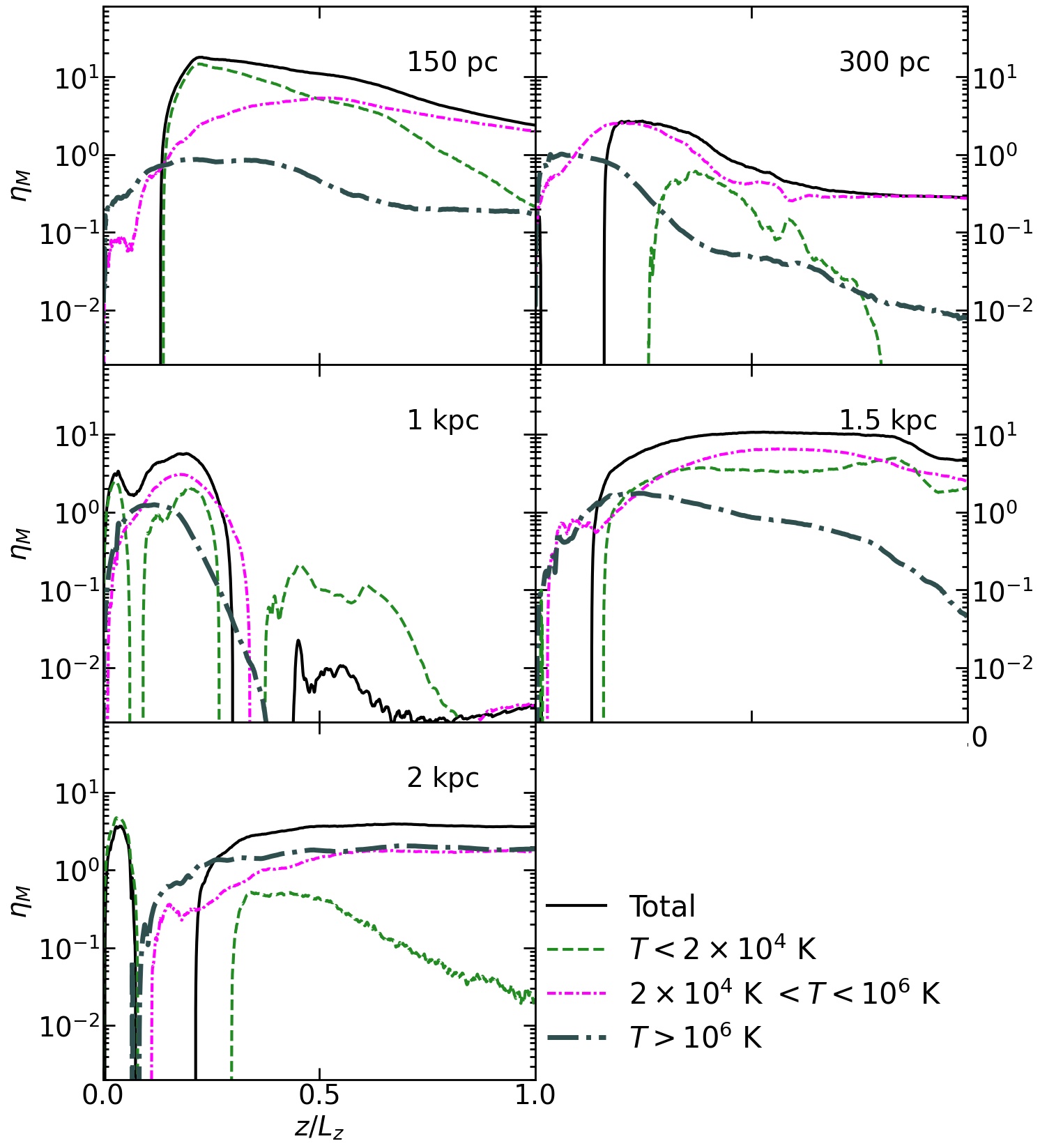}
    \caption{Mass-loading factor for $\Sigma2.5$-Z1-Hhh series. Labels in each panel is the value of $h_\mathrm{SN}$. The curves are time averages, with different colours representing different phases as indicated in the legend. }
    \label{fig:mass_loading_hSN}

\end{figure*}

Our experiments with $Z_{\rm bg}$ indicate that many properties of outflows, for example whether they are steady or bursty and their characteristic phase structure and loading factors, are determined by whether most SNe detonate in a dense or rare medium. To further explore this effect, we carry out the $\Sigma$2.5-Z1-Hhhh series of runs, which are identical in all respects except the scale height of SN injection, which takes on the values 150, 300, 1000, 1500, and 2000 pc. We remind the readers that the vertical half-height of the simulation domain in these runs is $L_z=\pm 8$ kpc to accommodate the larger scale height of the galaxy in weakened gravity due to the smaller gas and stellar surface density (see \autoref{tab:params}) and the correspondingly larger distance from the plane required for winds to attain their steady-state properties. 

Firstly, in \autoref{fig:gas_scale_height} we show how the gas scale height ($h_{\rm gas}$) varies as a function of time in runs with varying with $h_{\rm SN}$, where we define $h_{\rm gas}=(h_\mathrm{gas}^+ + h_\mathrm{gas}^-)/2$. Here $h_\mathrm{gas}^+$ is the scale height in the top half of the domain, which we define implicitly by the condition that $\int_0^{h_\mathrm{gas}} \int_0^L \int_0^L \rho \, dx\, dy\, dz = M_\mathrm{gas}/2e$, where $M_\mathrm{gas}$ is the total gas mass in the simulation domain, so that for an exponential density distribution $h_\mathrm{gas}^+$ reduces to the usual exponential scale height; 
our definition for the scale height in the lower half of the box, $h_\mathrm{gas}^-$, is analogous. We see from \autoref{fig:gas_scale_height} that the cases with $h_\mathrm{SN} = 150$ and 300 pc have $h_\mathrm{gas} \gg h_\mathrm{SN}$, the cases with $h_\mathrm{SN} = 1.5$ and 2 kpc have $h_\mathrm{gas}\ll h_\mathrm{SN}$, and that for $h_\mathrm{SN} = 1$ kpc we have $h_\mathrm{SN} \approx h_\mathrm{gas}$. For the first two cases the gas produced by SNe is efficiently trapped near the midplane, driving the initial gas away and yielding a large scale height, while for the latter two only a fraction of SNe occur mixed with or below the ISM, allowing the ISM to form a thinner disc near the midplane. The $h_\mathrm{SN}=1$ kpc case is intermediate between these two. Though we do not show this explicitly here, both the $\Sigma50$-runs and $\Sigma13$-Z1-H150 belong to the second category of having $h_\mathrm{gas}\ll h_\mathrm{SN}$ while $\Sigma13$-Z0.2-H150 falls in the $h_\mathrm{gas}\gg h_\mathrm{SN}$ category.

\autoref{fig:mass_loading_hSN} shows the time-averaged mass loading factor for the $\Sigma$2.5-Z1-Hhhh at the edge of the box, computed in the same way as for \autoref{fig:loading_factors}. Comparing the gas scale height and the mass-loading, we see a trend. For 150 and 300 pc the mass loading is dominated by cool and warm unstable gas. As noted earlier, for these cases $h_\mathrm{gas} \gg h_\mathrm{SN}$. For the $h_\mathrm{SN} = 1$ kpc case, while $h_\mathrm{gas} < h_\mathrm {SN}$ initially, we find that $h_\mathrm{gas}$ grows with time and this effectively stops mass outflow by the end of the simulation. For the $h_\mathrm{SN} = 1.5$ and 2 kpc cases, the SN decidedly go off at heights larger than the gas scale height and the outflow. In these two cases there is a steady outflow, with the former case dominated by cool and warm unstable gas and the latter by a mix of hot and warm unstable gas.

Consulting \autoref{tab:results}, we see that this variation with $h_\mathrm{SN}$ also shows up in energy and metal loading. The runs with $h_\mathrm{SN} < 1.5$ kpc all have low energy loading and little metal loss, $h_\mathrm{SN} = 1.5$ kpc has moderate energy loading and metal loss, and $h_\mathrm{SN}=2$ kpc has high energy loading and metal loss.

\section{Discussion}\label{sec:disscussion}

Here we seek to draw some overall lessons from our grid of models. We first define some broad classifications of winds in \autoref{subsec:three_types}, and we discuss the implications of this classification for metal loading and galactic chemical evolution in \autoref{subsec:implications}. We also make some observations on the importance of boundary conditions and compare with previous work in \autoref{subsec:boundary}.

\subsection{The three types of winds}
\label{subsec:three_types}

\begin{figure}
    \includegraphics[width=\columnwidth]{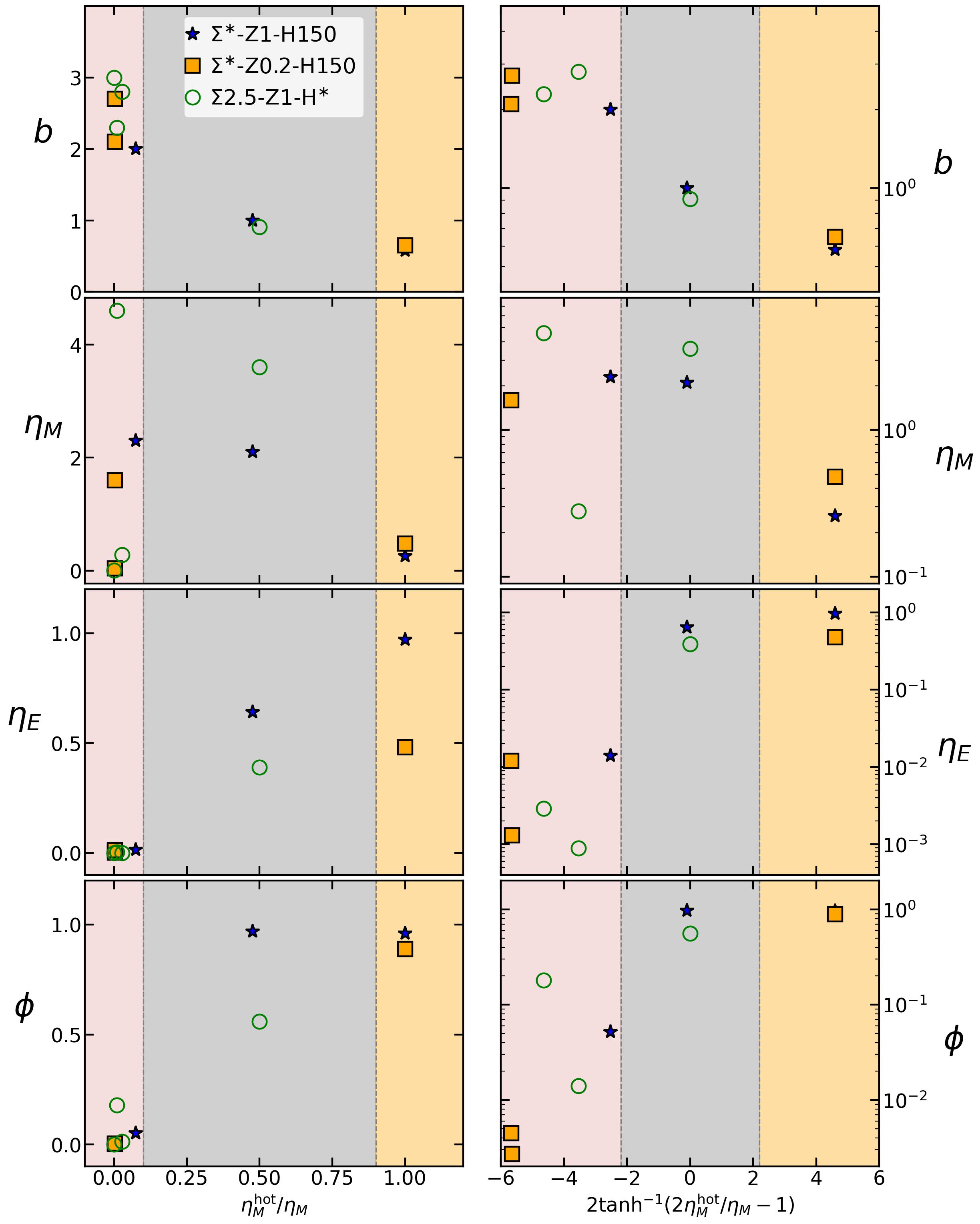}
    \caption{
    Burstiness parameter $b$, mass loading factor $\eta_M$, energy loading factor $\eta_E$, and fraction of metals lost to the wind $\phi$ (top to bottom) as a function of $\eta_M^\mathrm{hot}/\eta_M$, the fraction of wind mass flux in the hot ($T>10^6$ K) phase. Different symbols indicate different run series, as indicated in the legend. The left column shows results on a linear scale, while the right column has a logarithmic $y$ axis and shows the quantity $2\tanh^{-1}(2\eta_M^\mathrm{hot}/\eta_M-1)$ on the $x$ axis, which approaches a logarithmic scaling both as $\eta_M^\mathrm{hot}/\eta_M\to 0$ and $\eta_M^\mathrm{hot}/\eta_M\to 1$.
    \label{fig:windtypes}
    }
\end{figure}

The galactic winds we have found can be divided into three broad categories: cool and bursty winds with low energy and metal loading, hot and steady winds with high energy and metal loading but low mass loading, and multiphase winds with intermediate energy loading, burstiness, and mass loading. We illustrate these three categories in \autoref{fig:windtypes}, where we show $b$, $\eta_M$, $\eta_E$, and $\phi$ as a function of $\eta_M^\mathrm{hot}/\eta_M$, the fraction of the total outflow mass flux in the hot phase. As the plot shows, our runs tend to fall into three groups of hot mass flux fraction -- low ($\eta_M^\mathrm{hot}/\eta_M \lesssim 0.1$), medium ($0.1 \lesssim \eta_M^\mathrm{hot}/\eta_M \lesssim 0.9$), and high ($\eta_M^\mathrm{hot}/\eta_M \gtrsim 0.9$) -- highlighted with pink, grey, and yellow bands in the figure. We emphasise that these divisions are rough, and because our sampling of parameter space is coarse we have limited ability to distinguish sharp transitions between regimes from smooth variation. Nonetheless, this categorisation is useful, because we see that all of the parameters describing the wind are varying systematically between these groups, albeit with a fair amount of scatter within groups. 

Variation with $\eta_M^\mathrm{hot}/\eta_M$ -- and evidence for three distinct groups -- is particularly clear for $b$, $\eta_E$, and $\phi$. For these quantities we see that cool winds with $\eta_M^\mathrm{hot}/\eta_M \lesssim 0.1$ have distinctly higher burstiness and lower energy and metal loading than multiphase or hot winds, but that there is no strong dependence on $\eta_M^\mathrm{hot}/\eta_M$ once this value is below $\sim 0.1$. For mass loading $\eta_M$ the situation is slightly less clear, as runs with cool winds appear to be capable of a range of mass loadings, but there remains a clear difference between hot and multiphase winds. The correlation between wind temperature and energy and metal-loading has been noted before by \citet{Li&Bryan20}, and effectively determines the mode of feedback in galaxies -- preventative or ejective \citep{Carr+23}. High energy loading coupled with low mass loading, the typical outcome in the yellow band, will heat up the CGM inhibiting gas cooling, leading to preventative feedback. Runs within the pink band will be ejective in nature. 

Based on our results, which family a given galaxy falls into seems to depend primarily in the scale height of supernovae as compared to that of the warm ISM: cases where the SNe are confined closer to the midplane than the warm ISM produce cool and bursty winds, those where the SNe are more extended than the ISM produce hot winds, and those where the scale heights are similar appear to favour multiphase winds. Of course these immediately invites the question of how this ratio varies in real galaxies. One might at first expect the multiphase case to be most prevalent, since one might naively expect that star formation should follow the same vertical distribution as ISM mass, and that in turn this would guarantee that SNe have roughly the same scale height as the ISM. However, this ignores several potential complications.

Two effects that tend to favour large SN scale heights are runaways and type Ia SNe. With regard to the first of these, roughly $30\%$ of Milky Way O stars are observed to be runaways \citep[as well as earlier references therein]{Carretero-Castrillo23a} with velocities of tens of km s$^{-1}$ relative to the mean of the Galactic plane, enough for the more long-lived among them ($\approx 30-40$ Myr lifetime prior to explosion) to travel an appreciable fraction of a kpc. This population will produce at least a tail of SN explosions well above the vertical zone in a galactic disc where stars form leading to stronger outflows with larger loading factors \citep{Andersson+20, Steinwandel+23}. Similarly, in galaxies with older stellar populations such as the Milky Way, half of SNe are type Ia rather than type II \citep[e.g.,][]{Ruiter09a, Adams13a}, and the scale height of these explosions is typically $2-3\times$ larger than that of type II SNe \citep{Hakobyan17a}.

Conversely, the naive assumption that the scale height of star formation should match that of the ISM ignores the fact that the ISM contains multiple distinct phases, and that both observations \citep[e.g.,][]{Bigiel08a, Leroy08a} and theory \citep{Krumholz11b, Glover11b} agree that star formation correlates strongly only with the molecular phase. There is substantial evidence, in turn, that the molecular phase \citep{Jeffreson22a} and the cold atomic phase that traces it closely \citep{Dickey22a, McClure-Griffiths23a} have a smaller scale height than the warm atomic phase, at least in galaxies like the Milky Way. 

Since we have found that the properties of the wind can be sensitive to even relatively small changes in the scale height of the SNe (for example runs $\Sigma$2.5-Z1-H1000 versus H1500 and H2000), the actual regime into which any given galaxy falls may well depend on details such as the relative scale heights of the different ISM phases, the frequency and velocity distribution of runaways, and the relative frequencies of type II and type Ia SNe. This in turn is a substantial challenge for numerical simulations of galactic wind launching. The implication of our finding is that a complete treatment of the problem requires not only resolving the separate phases of the ISM, but also resolving their separate scale heights, and at a minimum including models for the stellar dynamical processes that can lead to differences between the vertical distributions of star formation and stellar explosions.

\subsection{Implication for galactic chemical evolution}
\label{subsec:implications}

The sensitivity of wind metal loading to the relative scale heights of the gas and SNe has potentially important implications for galactic chemical evolution. In our numerical experiments here we have considered only the simple case of a single population of SN progenitors with a single scale height, but this is of course an oversimplification. It is therefore interesting to speculate how our results might generalise to the more realistic case of multiple metal injection sites.

Galaxies have four main nucleosynthetic sites: type II SNe as we have considered here, type Ia SNe, AGB stars, and neutron star mergers. Each of these populations may be extended over different scale heights depending on the galaxy's star formation history. In general older populations will have larger scale heights due to the well-known stellar age-velocity dispersion correlation \citep[e.g.,][]{Nordstrom04a, Holmberg09a}, so SNII that immediately follow star formation will be in the thinnest disc (with the exception of runaways), followed by SNIa -- which are observed to have scale heights roughly twice that of SNII \citep{Hakobyan17a} -- and AGB stars (whose relative heights likely depend on the exact mass range within the AGB star population on which we focus), and neutron star mergers likely farthest out both due to their long delay times and the asymmetric kicks that likely accompany these systems' birth. Given that we find that metal loss from a population depends on the height at which it injects its metals, it seems likely that not only is $\phi$ non-zero for all of these populations, contrary to the assumption most commonly made in chemical evolution modeling (for a recent example see \citealt{Kravtsov22a}, whose framework allows for the possibility of metal loading, but who choose $\phi=0$ as their default value), but that \textit{it is potentially different for each of them}, likely in the direction of greater metal loss for nucleosynthetic injection by the oldest populations, which have had the longest time to undergo dynamical heating.
This complicates chemical modelling substantially, and muddies interpretations based on the simple assumption of no, or uniform, metal loading.

As an example of a possible complications introduced by differential metal loading, consider the $\alpha$-Fe ratios of early-type galaxies (ETGs). The atmospheres of ETG stars have a higher ratio of $\alpha$ elements (O, Mg, etc) to Fe than is found in the Sun \citep{Peterson76, Worthey+92, Milone+00}, and this is traditionally interpreted as indicating very rapid star formation in these systems, such that many of the stars formed before SNIa, with their longer delay times, were able to enrich the ISM. While this is certainly consistent with the data, our results offer a possible alternate physical explanation of how $\alpha$-Fe enrichment might occur: through differential metal loading of metals injected at different heights. In this scenario, the yields of SNIa -- most prominently Fe -- would be lost more easily than the $\alpha$-elements, which come primarily from SNII. If differential metal loading were more important in the progenitors of ETGs than in the progenitors of lower-mass galaxies such as the Milky Way, that would produce much the same signature as extremely rapid star formation that ran to completion on timescales smaller than the SNIa delay time. Determining the extent to which this effect might have contributed to the $\alpha$/Fe differences between ETGs and lower-mass galaxies will require a more quantitative exploration of the relationship between galaxy scale height differences and variations in \fy. We are currently running a set of simulations to address this questions, and will report the results in a forthcoming paper.

\subsection{On the importance of vertical boundary conditions in tall box simulations}
\label{subsec:boundary}

A final lesson to be drawn from our numerical experiments is the importance of boundary conditions at the $\pm z$ edges of the computational domain in tall-box experiments such as ours. As noted in \autoref{sec:methods}, in \qedi~we used inflow/outflow boundary conditions in the vertical direction, which are implemented as a first-order extrapolation of the conditions inside the computational domain into the ghost zones. However, we find that in at least some cases using this boundary condition over long times leads to a slow-moving inflow of low-density, cool gas into the domain, which mixes with and modifies the properties of the outflow. This condition is triggered when there is dense, cool gas near the boundary with a velocity back inwards towards the plane, as happens for example with fountain flows where cool gas is driven upwards fast enough to reach the edge of the domain but not fully escape. When this happens, applying first order extrapolation at the boundary produces an additional supply of cool gas entering the domain, and this becomes self-reinforcing: the boundary creates new inflowing gas, which increases the inflow momentum and makes the inflow harder to turn around, which in turn generates yet more inflow.

Similar behaviour has also been reported elsewhere in the literature. For example \cite{Caproni+23} show that ``open'' boundary conditions (equivalent to our inflow/outflow conditions) act as reservoirs of gas flowing into the domain in their simulations. Similarly, \citet{Melso+19} find that even when the inflow is warm rather than cool, this can still contribute to the production of cold gas at large heights, because a diffuse warm inflow can become Rayleigh-Taylor and Kelvin-Helmholtz unstable when it shocks against a fast, hot, SN-driven outflow. The resulting instabilities lead to the production of cold ($T<10^3$ K) clouds, with the total mass of cold gas produced linked to the relative speed of the inflowing and the outflowing gas.

As explained in \autoref{sec:methods}, we avoid these effects by using ``diode'' boundaries \citep{Fryxell+00, Zingale+02, Caproni+23} that allow gas to flow out of the domain but not back in. The fact that we do so may play a significant role in explaining why our results differ from those of the SMAUG suite \citep{Kim+20}, which uses inflow/outflow as we did in \qedi. Our $\Sigma$50-Z1-H150 run is quite similar to their R4 case, but \citeauthor{Kim+20} find that their outflows in this case are dominated by the cool phase ($T<2\times 10^4$ K), which produces a mass loading an order of magnitude larger than the other phases. This phase also carries the majority of the metal flux. By contrast, our $\Sigma$50-Z1-H150 run is hot-gas dominated in both mass and metal flux. There are several possible contributors to this discrepancy other than boundary conditions: our runs differ in the size of the domain and in particular the amount of distance above and below the plane that they include ($1\times 1\times 8$ kpc for us, $0.5\times 0.5\times 4$ kpc for them), the treatment of gas self-gravity (fixed potential for us, computed on-the-fly for them), and the star formation and feedback recipe (fixed scale height for us, self-consistent for them). However, there is circumstantial evidence that boundary conditions are an important contributor: examining \citeauthor{Kim+20}'s Figures 1 and 3, it is clear that, for at least a part of the run duration, their simulation is experiencing considerable inflow from the edge of the domain, providing a supply of cool at one epoch that can then be pushed out at a later epoch and thereby raising to cool contribution to the mass loading factor. Such inflows are expressly forbidden by our boundary conditions, suggesting that the boundary conditions may be a substantial contributor to the differences between our results and theirs.

We emphasise that neither choice of boundary condition is perfect: while the inflow/outflow conditions used in SMAUG allow artificial generation of cool gas at the top of the domain, ours perhaps suffer from the opposite problem of artificial deletion of cool gas, since we assume that everything that reaches $\pm 4$ kpc is able to escape to infinity, whereas in reality presumably some of this gas does fall back as a fountain flow. Moreover, both the SMAUG boundary conditions and ours -- in regions where the velocity vector is pointing outward and we too use inflow/outlow -- may suffer from a defect generically observed with extrapolating outflow boundary conditions that arises because the flow of information along characteristics is not precisely controlled. When the outflow is subsonic, such boundary conditions have been shown to exhibit unphysical behavior, for example, causing vortices near the boundary to be ``vacuumed'' out of the domain \citep{Motheau_2017}. This appears to be an issue with all subsonic outflow boundary conditions currently used in the astrophysical literature.

A broader lesson to be drawn from this discussion is that, in cases such as our $\Sigma$50-Z1-H150 run where the choice of $z$-boundary condition (and by extension the vertical box size) appears to affect the outcome significantly, the tall box paradigm may simply be inadequate to the problem. As several authors have noted \citep[e.g.,][and references therein]{Thompson24a}, the physics of galactic winds is largely controlled by the development of a sonic point that separates the launching region close to the galaxy from the wind region, and prevents the back-propagation of information from the wind region into the galaxy. In the tall box approximation, however, there is no sonic point, because streamlines do not diverge and the escape speed is formally infinite. The ability of the $z$-boundary to affect the wind launching region close to the galaxy in both our simulations and in the SMAUG suite is a direct consequence of this lack of a sonic point. The fact that this seems to matter to the qualitative outcome means we may simply be reaching the limits of what tall box simulations can tell us about galactic winds.

The obvious question raised by this discussion is: what is the alternative? One possibility is to use simulations of entire isolated galaxies \citep[e.g.,][]{Rey24a, Steinwandel+24}. This approach removes the problem of the missing sonic point, but at the price of requiring the inclusion of some sort of model for the halo into which the outflow propagates, and the possibility that mixing between the outflow and the halo might artificially alter the outflow properties. Moreover, while the resolution of such isolated galaxy simulations is considerably better than can be achieved in cosmological simulations, and may well be sufficient to study mass or energy loading, it is likely not sufficient to for metal loading. For example, \citeauthor{Rey24a} reach a peak resolution of 18 pc, while \citeauthor{Steinwandel+24} use a Lagrangian method with a mass resolution of 4 M$_\odot$, which for the $n\sim 10^{-3}$ cm$^{-3}$ densities characteristic of the hot gas that carries most of the metals (c.f.~Figure 8 of \citetalias{QEDI}) corresponds to a linear resolution of roughly 50 pc. By contrast, in \citetalias{QEDI} we found that metal outflow rates and loading factors do not converge until $\approx 2$ pc resolution. Fully cosmological simulations avoid the problem of needing an artificial halo, but at the cost of making the resolution problem even worse.

Given this situation, one possible way forward would be to run isolated-galaxy or cosmological simulations, but then use adaptive techniques to zoom in on isolated sections of galactic discs to reach the resolutions required to study metal loading. In the isolated-galaxy case, this effort could involve multiple possible background halo models to control for their uncertain effects. This seems like the most promising approach given the problems that exist with each of the primary methods currently in use.

\section{Summary and Conclusions}

We present results from the QED simulations, a suite of tall box simulations of galactic wind launching and metal loading based on \textsc{Quokka} \citep{Wibking23a, He24a}, a new code that leverages GPU acceleration to allow simulation volumes and resolutions substantially larger than possible using earlier CPU-based codes. The QED simulations evolve a patch of star-forming galaxy to understand the properties of outflows generated by supernova feedback within it, and feature high and uniform resolution throughout a large volume so that we can resolve the different gas phases and the exchange of mass, metals, and energy between them out to distances far off the galactic disc. We conduct 10 simulations to explore a parameter space defined by three axes: gas surface density, which sets the both the depth of the gravitational potential well and the supernova rate, metallicity which affects the gas rate at which gas cools, and the scale height at which supernovae occur, which determines the fraction of explosions that occur in the dense midplane versus above it. We evolve all simulations for $\gtrsim 200$ Myr, long enough for outflows to reach statistical steady state, and then we measure spatial and temporal averages of the mass, energy, and metal outflow rates and loading factors. We also estimate \fy, which quantifies the fraction of SN-injected metals that are promptly lost to outflows rather than being retained in the disc. 

Our major conclusions are as follows:

\begin{enumerate}
    
    \item \textbf{Three types of winds.} We find that outflows can be sorted into three major categories. ``Steady and hot'' outflows (for example the $\Sigma50$ runs) are dominated by hot gas and characterised by weak burstiness and small mass loading but metal and energy loading, with most of the metals lost promptly to outflow. For ``cool and bursty'' winds the hot phase is subdominant and the cooler phase sets the outflow properties; mass loading is moderate to high, but energy and metal loading are small, and most metals are retained. The third type, ``multiphase'', are characterised by outflows with significant contributions from both cold and hot phases, moderate levels of burstiness, and intermediate loading factors and metal loss. 
       
    \item \textbf{Ratio of supernova and gas scale heights is a key parameter.} The most important factor that determines into which of these categories a given outflow will fall is where exactly the SN go off relative to the gas. If SNe occur predominantly in a low density environment above the dense gas ($h_{\rm SN}>h_{\rm gas}$), we find the hot and steady regime, while if they occur primarily in a high density location within the dense gas layer ($h_{\rm SN}<h_{\rm gas}$), we find the cool and bursty regime. When the SN scale height is intermediate, $h_\mathrm{SN} \sim h_\mathrm{gas}$, the multiphase regime predominates.

    \item \textbf{Gas cooling affects the SN to gas scale height ratio.} Systems with sub-Solar metallicity tend to have larger effective gas scale heights, because the background ISM does not rapidly break up into cool clumps, and instead is more likely to exist in a pressure-supported warm case with larger scale height. This in turn leads to the somewhat paradoxical outcome that lower-metallicity systems may on average have cooler winds, because the lack of cooling in the background ISM makes it harder for SN-heated gas to break out. Instead, this gas can be trapped near the midplane, limiting outflow break out.
\end{enumerate}

One important implication of our work is that any mechanisms capable of altering the distribution of supernovae relative to the ISM -- including runaways and type Ia SNe, which increase the SN scale height relative to the gas scale height, and preferential star formation in cold ISM phases confined closer to the midplane, which tend to decrease it -- can have potentially strong effects on the nature of outflows. Even numerical treatments that include a self-consistent model for star formation based on gas self-gravity may not produce reliable results if they do not properly include these effects. In future work with QED we intend to consider self-consistent star formation models, but based on this finding it is clear that realistic treatments of ISM phase structure and models for runaways and type Ia SNe must be included as well.

A second implication is that, in a realistic galaxy with different nucleosynthetic sources operating at different scale heights, the prompt metal loss rate may vary significantly from one source to another. This complicates interpretations of chemical markers such as the $\alpha$/Fe ratio, since there is a degeneracy between the effects of differential metal loss and differences in delay times. Future chemical evolution modeling will need to consider how to cope with this degeneracy.

\section*{Acknowledgements}
AV and MRK acknowledge support from the Australian Research Council through awards FL220100020 and DP230101055. 
This work was supported by resources provided by the National Computational Infrastructure (NCI Australia), an NCRIS enabled capability supported by the Australian Government, the Pawsey Supercomputing Research Centre’s Setonix Supercomputer (https://doi.org/10.48569/18sb-8s43), with funding from the Australian Government and the Government of Western Australia, and the Oak Ridge Leadership Computing Facility at the Oak Ridge National Laboratory, which is supported by the Office of Science of the U.S. Department of Energy under Contract No. DE-AC05-00OR22725. We thank the referee, Martin Rey, for his insightful comments on the manuscript.

%%%%%%%%%%%%%%%%%%%%%%%%%%%%%%%%%%%%%%%%%%%%%%%%%%
\section*{Data Availability}

The initial condition files used to generate all the simulations in this work are available in the \textsc{Quokka} repository at \url{https://github.com/quokka-astro/quokka}. 
The raw simulation outputs are not provided due to their large size, but will be shared on reasonable request. Movies from the QED simulations can be found at \url{https://quokka-astro.github.io/quokka/gallery/}.

\section*{Software}

This work used the following software: \textsc{Quokka} \citep[\url{https://github.com/quokka-astro/quokka}]{Wibking23a, He24a}, \textsc{AMReX} \citep[\url{https://github.com/AMReX-Codes/amrex}]{AMReX_JOSS}, \textsc{yt} \citep[\url{https://yt-project.org/}]{Turk11a}, \textsc{numpy} \citep[\url{https://numpy.org}]{numpy}, and \textsc{matplotlib} \citep[\url{https://matplotlib.org/}]{matplotlib}.

%%%%%%%%%%%%%%%%%%%% REFERENCES %%%%%%%%%%%%%%%%%%

% The best way to enter references is to use BibTeX:

\bibliographystyle{mnras}
% \bibliography{references} % if your bibtex file is called example.bib

% Alternatively you could enter them by hand, like this:
% This method is tedious and prone to error if you have lots of references
%\begin{thebibliography}{99}
%\bibitem[\protect\citeauthoryear{Author}{2012}]{Author2012}
%Author A.~N., 2013, Journal of Improbable Astronomy, 1, 1
%\bibitem[\protect\citeauthoryear{Others}{2013}]{Others2013}
%Others S., 2012, Journal of Interesting Stuff, 17, 198
%\end{thebibliography}

%%%%%%%%%%%%%%%%%%%%%%%%%%%%%%%%%%%%%%%%%%%%%%%%%%

%%%%%%%%%%%%%%%%% APPENDICES %%%%%%%%%%%%%%%%%%%%%

\appendix
\section{Implications of neglecting local metallicity variations for gas cooling rates}\label{app:cooling}

\begin{figure}
    \includegraphics[width=\columnwidth]{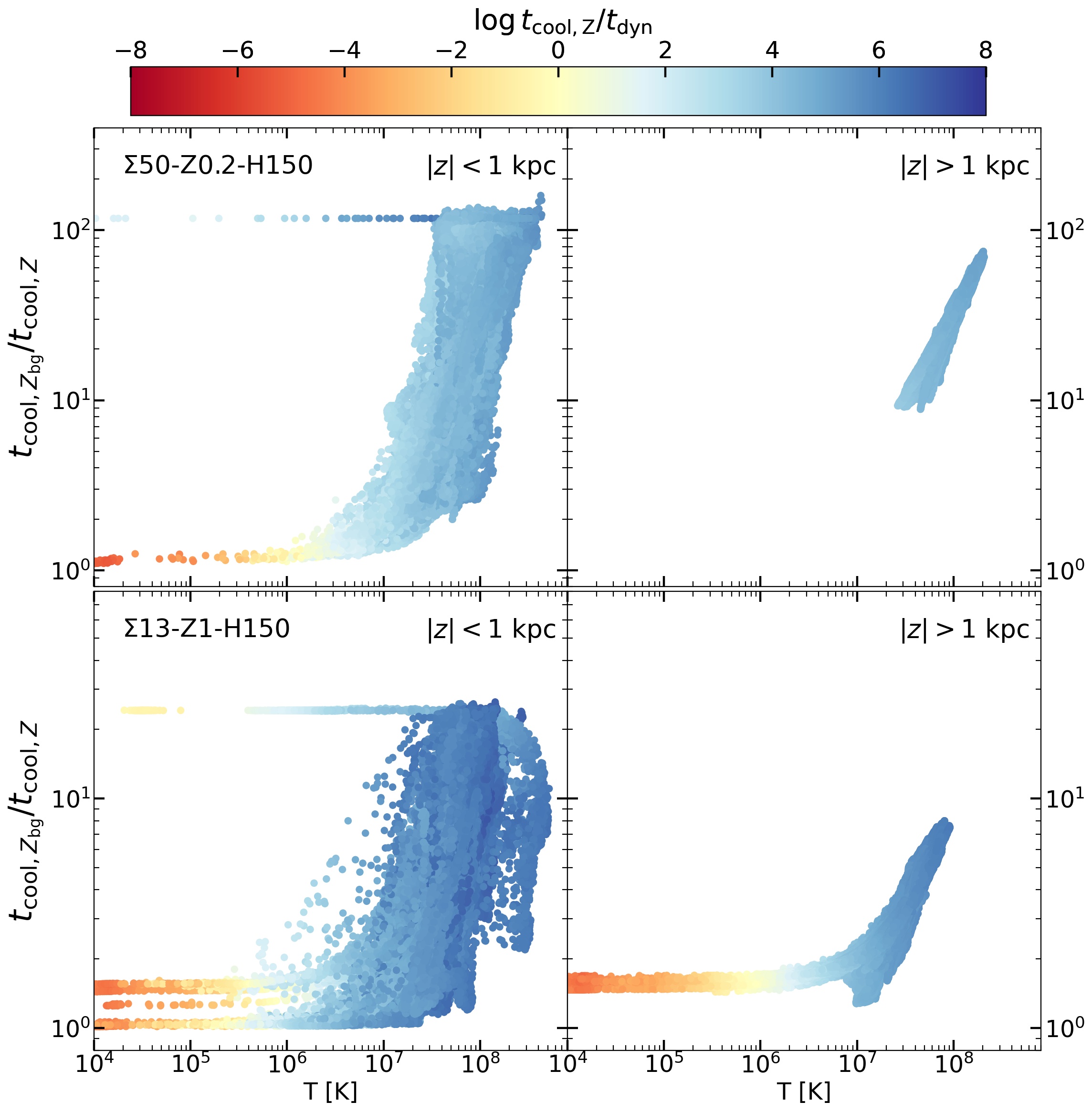}
    \caption{
    The distribution of the ratio of the two cooling times - $t_{\rm cool, Z_{\odot}}$ and $t_{\rm cool, Z}$ - with temperature for $\Sigma50$-Z0.2-H150 (top) and $\Sigma13$-Z1-H150 (bottom). While $t_{\rm cool, Z_{\rm bg}}$ is cooling derived assuming a constant background metallicity (0.2 and 1), $t_{\rm cool, Z}$ is taking into account the metal enrichment from SNe. The color coding represents $t_{\rm cool, Z}/t_{\rm dyn}$. Though gas hotter than $10^7$ K is enriched much higher than solar, its cooling time is much larger than its dynamical time.
    %This figure is just for one slice. Will be updated soon.
    }
    \label{fig:tcool}
    
\end{figure}

As discussed in \autoref{subsec:sn_metal_cooling}, the metal cooling rate in the QED simulations is set by the initial metallicity and does not change in response to local metallicity variations. We use this approach in order to be able to conduct clean experiments without having to worry about the effects of self-enrichment, but it means that we ignore the enhancements in cooling that should occur in highly metal-enriched gas containing large amounts of fresh SN ejecta. In this appendix we investigate the implications of this choice. To do so, we choose one snapshot each from the $\Sigma50$-Z0.2-H150 and the $\Sigma13$-Z1-H150 run as examples, and for the midplane slices in these snapshots we estimate two cooling times: $t_{\rm cool, Z_{\rm bg}}$ and $t_{\rm cool, Z}$. $t_{\rm cool, Z_{\rm bg}}$ is the cooling time of a gas parcel assuming it has metallicity identical to the background (that is 0.2$Z_{\odot}$ for the $\Sigma50$ case and $Z_{\odot}$ for the $\Sigma13$ case), which is the cooling prescription we use in the simulation. $t_{\rm cool, Z}$ is the cooling time if we were instead to use the true metallicity, $Z = \rho_Z /\rho$ which includes contribution from the background at $Z_{\rm bg}$ and SN enrichment. In both cases we define the cooling time as the ratio of the gas thermal energy per unit volume to the cooling rate per unit volume. The ratio $t_{\rm cool, Z_{\odot}}/t_{\rm cool, Z}$ quantifies the amount by which we overestimate the cooling time by neglecting local metal injection.

We plot the ratio against the gas temperature in \autoref{fig:tcool}; in this plot, every dot represents a cell. The two columns show gas in the disc ($|z|<1$ kpc) and outflow ($|z|>1$ kpc) regions. We separate these two because the ejecta are least mixed with the background gas in the disc region, and this is therefore where we expect the ratio $t_{\rm cool, Z_{\odot}}/t_{\rm cool, Z}$ to be largest. As the plot shows, our neglect of local metal enhancement has minimal effects in gas with  temperatures $\lesssim 5\times 10^6$ K; including local enhancement would make only a tens of percent difference for this material. The situation is quite different for gas at $T>10^7$ K, where the gas is more metal enriched. Here our metallicity assumption leads us to underestimate the cooling rate by factors of several in the wind region, and by up to order of magnitude in the disc.

However, it is now important to ask a follow-up question: does this matter to the dynamics? A large departure from the actual cooling time is only problematic if the cooling time itself is comparable to or shorter than the dynamical time -- to take an extreme example, if the true cooling time were 1 Gyr and we instead estimated it as 10 Gyr, this would have no effects on the dynamics because both times are so long that the gas would effectively behave as adiabatic over the duration of our simulation. For the purposes of evaluating where our simulations sit relative to this consideration, we define the dynamical time as $t_{\rm dyn}=L_z/c_s$, where $c_s$ is the sound speed of the gas. This is roughly the time required for hot gas traveling at the sound speed to exit the computational domain, and if the true cooling time is much longer than this then it does not matter much if we overestimate the cooling time, because even if we used a more accurate estimate gas would still exit the computational domain long before having time to cool. Given this consideration, we color the points in \autoref{fig:tcool} by $t_{\rm cool, Z}/t_{\rm dyn}$. The key point to take away from this colouring is that the gas where our neglect of local metal enhancement has significant effects -- e.g., causing us to overestimate the cooling time by a factor of two or more -- is also gas where the cooling time is orders of magnitude larger than the dynamical time, and thus cooling is not an important process. We can therefore conclude that neglecting local enhancement of cooling rates by local metal enrichment has relatively modest effects. While this conclusion is strong once superbubbles break out from the disc and outflows are established, we should bear in mind that superbubble breakout also depends on the cooling function and the background metallicity will alter the dynamics of these initial bubbles.

%%%%%%%%%%%%%%%%%%%%%%%%%%%%%%%%%%%%%%%%%%%%%%%%%%

% Don't change these lines
\bsp	% typesetting comment
\label{lastpage}
\end{document}